\documentclass[transmag]{IEEEtran}
\usepackage{subfigure}

\usepackage{cite}

\ifCLASSINFOpdf
  \usepackage[pdftex]{graphicx}
\else
\fi
\usepackage{amsmath}
\usepackage{array}
\usepackage{bm}
\usepackage{amssymb}
\usepackage{tikz}
\usetikzlibrary{patterns}
\usetikzlibrary{calc}
\usetikzlibrary{tikzmark,fit}
\usepackage{datatool,filecontents}

\DTLloaddb[noheader=false]{coordinates}{coordhex.dat}
\DTLloaddb[noheader=false]{coordinatesgreen}{coordhexgreen.dat}

\def\centerarc[#1](#2)(#3:#4:#5)
{ \draw[#1] ($(#2)+({#5*cos(#3)},{#5*sin(#3)})$) arc (#3:#4:#5); }
\tikzset{user/.style={circle,draw,inner sep=0pt, minimum size=0.1\R,},}
\tikzset{
	bs/.pic = {		
		\draw (0,0.4\R) -- (-0.2\R,-0.2\R);
		\draw (0,0.4\R) -- (0.2\R,-0.2\R);
		\draw (-0.2\R,-0.2\R) -- (0.133\R,0);
		\draw (0.2\R,-0.2\R) -- (-0.133\R,0);
		\draw (-0.133\R,0) -- (0.067\R,0.2\R);
		\draw (0.133\R,0) -- (-0.067\R,0.2\R);
		
		\draw (-0.21\R,0.4\R) -- (0.21\R,0.4\R);
		\draw \foreach \x in {-0.21, -0.14,...,0.22} {(\x\R,0.4\R) -- (\x\R,0.47\R)};
	}
}

\begin{document}
%
\title{Bare Demo of IEEEtran.cls\\ for IEEE Journals}
\title{Implementation of Machine Type Communication Using Non-Negative Least Squares}
\title{Implementation of Pilot-Hopping Sequence Detection for Grant-Free Random Access in Massive MIMO Systems}
\title{An Architecture for Pilot-Hopping Sequence Detection in Massive MIMO Systems with Grant-Free Random Access}
\title{Pilot-Hopping Sequence Detection Implementation for Grant-Free Random Access using Massive MIMO}
\title{Massive MIMO Pilot-Hopping Sequence Detection Architectures Based on Non-Negative Least Squares for Grant-Free Random Access}
\title{Massive Machine Type Communication Pilot-Hopping Sequence Detection Architectures Based on Non-Negative Least Squares for Grant-Free Random Access}
%
%
%

\author{Narges~Mohammadi~Sarband,~\IEEEmembership{Student~Member,~IEEE,}
	 Ema~Becirovic,~\IEEEmembership{Student~Member,~IEEE,}
	 Mattias~Krysander,\\
	 Erik~G.~Larsson,~\IEEEmembership{Fellow,~IEEE,}
	 and Oscar Gustafsson,~\IEEEmembership{Senior~Member,~IEEE}
\thanks{Manuscript received \today. This work was supported in part by the ELLIIT strategic research environment.}
\thanks{The authors are with the Department of Electrical and Engineering, Link\"oping University, SE--581 83 Link\"oping, Sweden,
	e-mails: \{narges.mohammadi.sarband, ema.becirovic, mattias.krysander, erik.g.larsson, oscar.gustafsson\}@liu.se.}
}

%
%

\markboth{Accepted to IEEE Open Journal of Circuits and Systems}%
{Sarband \MakeLowercase{\textit{et al.}}:mMTC based on NNLS}
%



\IEEEtitleabstractindextext{\begin{abstract}
User activity detection in grant-free random access massive machine type communication (mMTC) using pilot-hopping sequences can be formulated as solving a non-negative least squares (NNLS) problem. In this work, two architectures using different algorithms to solve the NNLS problem is proposed. The algorithms are implemented using a fully parallel approach and fixed-point arithmetic, leading to high detection rates and low power consumption. The first algorithm, fast projected gradients, converges faster to the optimal value. The second algorithm, multiplicative updates, is partially implemented in the logarithmic domain, and provides a smaller chip area and lower power consumption. For a detection rate of about one million detections per second, the chip area for the fast algorithm is about 0.7~mm$^2$ compared to about 0.5~mm$^2$ for the multiplicative algorithm when implemented in a 28~nm FD-SOI standard cell process at 1~V power supply voltage. The energy consumption is about 300~nJ/detection for the fast projected gradient algorithm using 256 iterations, leading to a convergence close to the theoretical. With 128 iterations, about 250~nJ/detection is required, with a detection performance on par with 192 iterations of the multiplicative algorithm for which about 100~nJ/detection is required.
\end{abstract}

}

\maketitle

%

\IEEEpeerreviewmaketitle

\section{Introduction}
%
%
%
%
\IEEEPARstart{M}{assive} machine type communication (mMTC) is a core use case in 5G and later generations wireless communication systems, where a large number of users are expected to be served \cite{ituvision2015,Bockelmann2018,Mikhaylov2019,Dang2020,Saad2020,Chen2020}. These users are expected to intermittently send small amounts of data, primarily in the uplink, i.e., from the users to the base station. As the mMTC users send a limited amount data, the impact of overhead signaling such as random access or scheduling requests is large. Therefore, a grant-free random access scheme is beneficial for these types of users. In a grant-free random access scheme all the active users will transmit their data at the same time without waiting for a grant from the base station. To be able to be distinguished by the base station, the active users also need to transmit a unique identifier.

Grant-free random access with massive MIMO, i.e., base stations with a large number of antennas, has been widely studied in many papers \cite{deCarvalho2017,Chen2018,Liu2018a,Liu2018b,Senel2018,Haghighatshoar2018,Becirovic2019,Chen2019,Shao2019,Ahn2019,Chen2020}. Conventionally in massive MIMO, the users transmit a pilot in each coherence interval that is used by the base station to estimate the channel. Often these pilots are mutually orthogonal such that the channel estimates will be good. When a grant-free random access scheme is applied to massive MIMO, the same pilots are used to detect the active users. If the number of users exceeds the dimension (in samples) of the channel coherence interval, the users cannot be allocated mutually orthogonal pilots, which leads to pilot collisions. There are two fundamentally different approaches for solving this problem. In the first,  pilots are drawn at random and allowed to be non-orthogonal \cite{Liu2018a,Chen2018,Senel2018,Haghighatshoar2018,Chen2019,Shao2019,Ahn2019}. In the other, the pilots are orthogonal, but the data transmission is spread over many coherence intervals \cite{deCarvalho2017,Becirovic2019}, i.e., pilot-hopping sequences are detected instead of pilots. To reduce the complexity of the detection problem, a common approach for grant-free user detection with massive MIMO is to utilize the sparsity of the solution, i.e., that only a small subset of users are active at the same time. Hence, many of the existing solutions use methods developed in the field of compressed sensing \cite{Chen2018,Liu2018a,Liu2018b,Senel2018,Haghighatshoar2018,Becirovic2019,Ahn2019}.

The scenario for detecting pilot-hopping sequences for grant-free random access was introduced in \cite{deCarvalho2017}. However, in \cite{deCarvalho2017} perfect detection of the pilot-hopping sequences was assumed. In \cite{Becirovic2019}, it was suggested that the non-negative least-squares (NNLS) is an appropriate problem formulation for the pilot-hopping sequence detection problem. 

The active users transmit one pilot per coherence interval (the time-frequency interval in which the channel can be assumed to be constant), according to a predetermined sequence. The pilot-hopping sequences are unique for every user but more than one user can transmit the same pilot in a coherence interval. The pilots in each coherence interval are accompanied by an uplink data part. An example of this scheme can be found in Fig. \ref{fig:coherence-interval-RA-MaMi}, with three users, two pilots and a pilot-hopping sequence of length four.
\begin{figure}
	\centering
	\begin{tikzpicture}[every node/.style={rounded corners, rectangle,fill=white,inner sep=1pt,scale=0.8},scale=0.5]
	\node at (-1.3,0.5) {User 3};
	\node at (-1.3,1.5) {User 2};
	\node at (-1.3,2.5) {User 1};
	
	\draw [pattern=north west lines, pattern color=black]
	(0,0) rectangle node{$\bm{\phi}_1$} (1,1);
	\draw [fill=white] 		
	(1,0) rectangle node{$D_3(1)$}(3,1);
	
	\draw [pattern=dots, pattern color=black] 	
	(3,0) rectangle node{$\bm{\phi}_2$}(4,1);
	\draw [fill=white] 		
	(4,0) rectangle node{$D_3(2)$}(6,1);
	
	\draw [pattern=north west lines, pattern color=black]
	(6,0) rectangle node{$\bm{\phi}_1$}(7,1);
	\draw [fill=white] 		
	(7,0) rectangle node{$D_3(3)$}(9,1);
	
	\draw [pattern=north west lines, pattern color=black]
	(9,0) rectangle node{$\bm{\phi}_1$}(10,1);
	\draw [fill=white] 		
	(10,0)rectangle node{$D_3(4)$}(12,1);
	
	\draw [pattern=north west lines, pattern color=black]	
	(0,1) rectangle node{$\bm{\phi}_1$}(1,2);
	\draw [fill=white] 		
	(1,1) rectangle node{$D_2(1)$}(3,2);
	
	\draw [pattern=north west lines, pattern color=black]
	(3,1) rectangle node{$\bm{\phi}_1$}(4,2);
	\draw [fill=white] 		
	(4,1) rectangle node{$D_2(2)$}(6,2);
	
	\draw [pattern=dots, pattern color=black] 	
	(6,1) rectangle node{$\bm{\phi}_2$}(7,2);
	\draw [fill=white] 		
	(7,1) rectangle node{$D_2(3)$}(9,2);
	
	\draw [pattern=dots, pattern color=black] 		
	(9,1) rectangle node{$\bm{\phi}_2$}(10,2);
	\draw [fill=white] 		
	(10,1)rectangle node{$D_2(4)$}(12,2);
	
	\draw [pattern=dots, pattern color=black] 	
	(0,2) rectangle node{$\bm{\phi}_2$}(1,3);
	\draw [fill=white] 		
	(1,2) rectangle node{$D_1(1)$}(3,3);
	
	\draw [pattern=north west lines, pattern color=black]	
	(3,2) rectangle node{$\bm{\phi}_1$}(4,3);
	\draw [fill=white] 		
	(4,2) rectangle node{$D_1(2)$}(6,3);
	
	\draw [pattern=north west lines, pattern color=black]
	(6,2) rectangle node{$\bm{\phi}_1$}(7,3);
	\draw [fill=white] 		
	(7,2) rectangle node{$D_1(3)$}(9,3);
	
	\draw [pattern=dots, pattern color=black] 	
	(9,2) rectangle node{$\bm{\phi}_2$}(10,3);
	\draw [fill=white] 		
	(10,2)rectangle node{$D_1(4)$}(12,3);
	\end{tikzpicture}
	\caption{Three users transmitting pilot-hopping sequences of length four using two pilots, $\bm{\phi}_1$ and $\bm{\phi}_2$. Additionally, in coherence interval $t$ the user $k$ also transmits uplink data $D_k(t)$.  }
	\label{fig:coherence-interval-RA-MaMi}
\end{figure}

In this paper, two architectures for solving this problem using the approach in \cite{Becirovic2019} is proposed. To the best of the authors' knowledge, the only other implementations of similar detectors are \cite{Tran2019}, implementing the algorithm from \cite{Senel2018}, and \cite{Henriksson2020}, implementing the algorithm from \cite{Haghighatshoar2018}. However, these are based on the other principle: having non-orthogonal pilot sequences drawn from random sources.

A preliminary version of this work was presented in \cite{Sarband2020}. In addition to a more detailed presentation and more detailed results, an alternative NNLS algorithm is also considered here. The implementation of the alternative algorithm is shown to require smaller chip area and less energy, although it has slightly worse detection performance for hard\footnote{Unfavorable channel hardening and unfavorable propagation such that the linearization in (\ref{eq:asymptotic-limit}) is not a good approximation.} channel conditions. 

In the next section, the detection algorithm is reviewed, algorithms for solving NNLS problems discussed, and the general proposed architectures introduced. Then, in Section~\ref{sec:mainresults}, implementation results and decisions are presented for a considered scenario. Finally, some concluding remarks are given in Section~\ref{sec:conclusions}.

\section{Methods and Procedures}
\subsection{Algorithm to Detect Pilot-Hoppping Sequence}
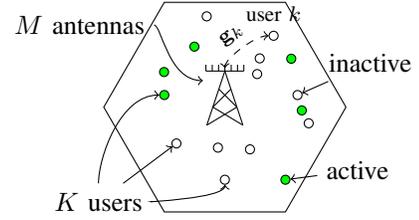
\begin{figure}
	\centering
	\begin{tikzpicture}[scale=0.7]
	\newdimen\R
	\R=1.2cm
	\newdimen\H
	\H=2.3cm
	\draw (0:\H) \foreach \x in {60,120,...,360} {  -- (\x:\H) };
	\node[user,fill=green] (k1) at (-0.5*\H,0.1*\H) {};
	\node[user] (k2) at (0*\H,-0.6*\H) {};
	\node[user] (k3) at (-0.4*\H,-0.3*\H) {};
	\node[user,label={\footnotesize user $k$}] (k) at (0.4025*\H,0.5898*\H) {};
	\path (-2\R,-1.5\R) node(K) {$K$ users};
	\draw[->] (K) .. controls +(up:0.7\R) and +(left:0.7\R) .. (k1);
	\draw[->] (K) .. controls +(right:0.7\R) and +(down:0.3\R)..(k2);
	\draw[->] (K)--(k3);
	\node[user,fill=green] (k4) at(0.5*\H,-0.6*\H) {};
	\node[user] (k5) at(0.6*\H,0.1*\H) {};
	\path (2.1\R,-\R) node(A) {active};
	\path (2.3\R,0.7\R) node(I) {inactive};
	\draw[->] (A)--(k4);
	\draw[->] (I)--(k5);
	\DTLforeach*{coordinates}{\x=x, \y=y}{
		\node[user] at (\x*\H,\y*\H) {};
	}
	\DTLforeach*{coordinatesgreen}{\x=x, \y=y}{
		\node[user,fill=green] at (\x*\H,\y*\H) {};
	}	
	\pic at (0,0) {bs};	
	\node (bs-point) at (-0.2\R,0.42\R) {};
	\node (antenna-point) at (0\R,0.47\R) {};
	\path (-2.3\R,1.3\R) node (M) {$M$ antennas};
	\draw[->] (M) .. controls +(right:1.3\R) and +(left:0.7\R) .. (bs-point);
	\draw[dashed,<->] (antenna-point) .. controls +(up:0.3\R) and +(left:0.3\R) .. node[above,sloped] {\footnotesize$\mathbf{g}_{k}$} (k) ;
	\end{tikzpicture}
	\caption{System model: a cell with a base station using $M$ antennas serving $K$ users, where only a subset is active. The channel between user $k$ and the base station is denoted by $\mathbf{g}_k$.}
	\label{fig:sm}
\end{figure}

We consider a single-cell massive MIMO system where the base station is equipped with $M$ antennas and serves $K$ mMTC users, each with a single antenna, see Fig. \ref{fig:sm}. We assume that only a subset of these users is active at once. We consider a block fading channel model where the channel of user $k$ in coherence interval $t$ is denoted by $\mathbf{g}_k^t \in \mathbb{C}^M$. The users transmit $T$ coherence interval long pilot-hopping sequences. There are $\tau_\text{p}$ mutually orthogonal pilots. Each pilot consists of  $\tau_\text{p}$ symbols and have unit norm. 
At the pilot phase of coherence interval $t$, the base station receives
\begin{equation}
\mathbf{Y}^t = \sum_{k=1}^{K}\sum_{j=1}^{\tau_\text{p}}\alpha_k S_{j,k}^t\sqrt{\tau_\text{p}p_k}\mathbf{g}_k^t\bm{\phi}_j^H + \mathbf{N}^t, 
\end{equation}
where
\begin{equation}
\alpha_k = \begin{cases}
1 & \text{if user }k\text{ is active,} \\
0 & \text{otherwise,}
\end{cases}
\end{equation}
\begin{equation}
S_{j,k}^t = \begin{cases}
1, & \text{if user }k\text{ sends pilot }j\text{ at pilot phase }t\text{,} \\
0, & \text{otherwise,}
\end{cases}\label{eq:Sdef}
\end{equation}
$p_k$ is the transmit power of user $k$, $\bm{\phi}_j \in \mathbb{C}^{\tau_\text{p}} $ is the $j$th pilot, and $\mathbf{N}^t \in \mathbb{C}^{M \times \tau_\text{p}}$ is noise with i.i.d. $\mathcal{CN}(0,\sigma^2)$ elements.
The model uses the assumption that the users start their pilot-hopping sequences at the same coherence interval. 

In each coherence interval the base station computes an estimate of the received signal energy over each pilot as 
\begin{equation}
E_{i,t} = \frac{(\mathbf{Y}^t\bm{\phi}_i)^H(\mathbf{Y}^t\bm{\phi}_i)}{M} - \sigma^2 = \frac{\lVert \mathbf{Y}^t\bm{\phi}_i \rVert^2}{M} - \sigma^2.
\end{equation}
With block Rayleigh fading,
\begin{equation}
\mathbf{g}_k\sim\mathcal{CN}\left(\mathbf{0},\beta_k\mathbf{I}_M\right),
\end{equation}
we have the channel hardening and favorable propagation properties of massive MIMO \cite{redbook}. The channel harding comes from 
\begin{equation}
\frac{\lVert\mathbf{g}\rVert^2}{M} \rightarrow \beta_k, \text{ as }M\rightarrow \infty,
\end{equation}
while the favorable propagation comes from
\begin{equation}
\frac{\mathbf{g}_k^H\mathbf{g}_{k'}}{M} \rightarrow 0,\text{ as }M\rightarrow\infty, k\neq{k'}.
\end{equation}
Using these properties, we can state the ``asymptotic energies'' as 
\begin{equation}
E_{i,t} = \frac{\lVert \mathbf{Y}^t\bm{\phi}_i \rVert^2}{M} - \sigma^2 \rightarrow \sum_{k=1}^{K}\alpha_k S_{i,k}^t\tau_\text{p}p_k\beta_k \text{ as }M\rightarrow \infty.
\label{eq:asymptotic-limit}
\end{equation}
From this limit, we can see that the asymptotic energies are linear in the user activity parameters. We assume that the users are using statistical channel inversion power control \cite{massivemimobook},
\begin{equation}
p_k = p \frac{\min_{k'}\beta_{k'}}{\beta_k} = p\frac{\beta_{\text{min}}}{\beta_k},
\end{equation}
where $\beta_{\text{min}} = \min_{k'} \beta_{k'}$ is the large-scale fading to the weakest user and $p$ is a system wide power parameter. With statistical channel information, the received signal power at the base station from each user is the same. 

We use the asymptotic energies and the fact that the users perform statistical channel inversion to form the column vector
\begin{equation}
\mathbf{b} = \frac{1}{\tau_\text{p}p\beta_{\text{min}}}
\begin{pmatrix}
E_{1,1} \\ \vdots \\ E_{\tau_\text{p},1} \\ \vdots \\ E_{i,t} \\ \vdots \\ E_{1,T}\\ \vdots \\ E_{\tau_\text{p},T}\label{eq:bvector}
\end{pmatrix}
\end{equation}
and the $T\tau_p \times K$-matrix 
\begin{equation}
\renewcommand\arraystretch{0.15}
\mathbf{A} = \begin{pmatrix}
S_{1,1}^1 & & \dots & &S_{1,K}^1 \\
\vdots & & \ddots & & \vdots \\
S_{\tau_\text{p},1}^1 & & \dots & &S_{\tau_\text{p},K}^1 \\
\vdots & & \ddots & &\vdots\\
S_{1,1}^T & & \dots & &S_{1,K}^T \\
\vdots & & \ddots & & \vdots \\
S_{\tau_\text{p},1}^T & & \dots & &S_{\tau_\text{p},K}^T \\
\end{pmatrix}.\label{eq:amatrix}
\end{equation}
One can note that $\sum_{j=1}^{\tau_p} S_{j, k}^t = 1,\ \forall k, t$, as user $k$ uses exactly one pilot sequence in coherence time interval $t$.

The $\mathbf{A}$-matrix for the three users in Fig.~\ref{fig:coherence-interval-RA-MaMi} is
\begin{equation}
\newcommand{\timezero}{%
	\left.\vphantom{\begin{matrix}0\\0\end{matrix}}\right\}%
	t = 1%
}
\newcommand{\timeone}{%
	\left.\vphantom{\begin{matrix}0\\0\end{matrix}}\right\}%
	t = 2%
}
\newcommand{\timetwo}{%
	\left.\vphantom{\begin{matrix}0\\0\end{matrix}}\right\}%
	t = 3%
}
\newcommand{\timethree}{%
	\left.\vphantom{\begin{matrix}0\\0\end{matrix}}\right\}%
	t = 4%
}
\mathbf{A} = 
\begin{matrix}
\text{User}
 & \\
 \begin{matrix}
 1 & 2 & 3 
 \end{matrix} & \\
\begin{pmatrix}\hspace*{0.2em}
\begin{matrix}
0 & 1 & 1\\
1 & 0 & 0
\end{matrix} \\
\hspace*{0.2em} \begin{matrix}
1 & \tikzmarknode{m1}{1} & 0\\
0 & \tikzmarknode{m2}{0} & 1\\
\end{matrix} \\
\hspace*{0.2em} \begin{matrix}
1 & 0 & 1\\
0 & 1 & 0\\
\end{matrix} \\
\hspace*{0.2em} \begin{matrix}
0 & 0 & 1\\
1 & 1 & 0
\end{matrix}
\end{pmatrix}
& \hspace*{-1em}\begin{matrix}
\timezero  \\ \timeone \\ \timetwo \\ \timethree
\end{matrix}
\end{matrix}
\begin{tikzpicture}[overlay,remember picture]
\node[fill=black,rounded corners,opacity=0.1,fit=(m1)(m2)]{};
\end{tikzpicture}\label{eq:aexample}
\end{equation}
The gray area corresponds to $S_{1,2}^2 = 1$ and $S_{2,2}^2 = 0$, i.e., user 2 transmits pilot sequence 1 in coherence interval 2.

With this vector and matrix, we can state the asymptotic energy equation as
\begin{equation}
\mathbf{Ax} \rightarrow \mathbf{b} \text{ as }M \rightarrow \infty,\label{eq:asymptotic-linear}
\end{equation}
where
$$\mathbf{x} = \begin{pmatrix}
\alpha_1 &\dots &\alpha_k &\dots &\alpha_K
\end{pmatrix}^T.$$

We use the fact that not all the users are active at the same time, i.e., that $\mathbf{x}$ is sparse, to solve the pilot-hopping sequence detection problem. In \cite{Becirovic2019}, it was shown that formulating this as an NNLS problem produce sparse solutions that solve the pilot-hopping sequence detection problem. It was also shown that using additional sparsity promoting techniques, such as NNLS combined with $L_1$-minimization, do not improve the results. 

\subsection{Algorithms for Solving NNLS}
The non-negative least squares (NNLS) problem can be stated as
\begin{equation}
\begin{array}{ll}
\text{minimize} & \| \mathbf{A}\mathbf{x} - \mathbf{b}\|_2 \\
\text{subject to} & \mathbf{x} \geq \mathbf{0},
\end{array}
\end{equation}
where the design matrix, $\mathbf{A}$ and an observation vector, $\mathbf{b}$, are given. $\mathbf{0}$ denotes a vector with all zeros. The solution vector, $\mathbf{x}$, minimizes the error of $\mathbf{A}\mathbf{x} - \mathbf{b}$ in the least squares sense subject to all elements in $\mathbf{x}$ being non-negative.
Over the years, many different algorithms have been proposed to solve NNLS problems. For an overview of different NNLS algorithms, we refer the reader to \cite{Slawski2013b}. 

The probably most well-known approach is the active set method by Lawson and Hanson \cite{Lawson1987}.
However, an active set-based algorithm relies heavily on branching making it unsuitable for a parallel high-speed implementation since parts of the execution will be sequential in nature.
Besides, the execution is data dependent making it non-deterministic. 

Instead, we here use two algorithms with a much higher degree of computational parallelism. As they are iterative, there are still limitations, but as later seen, a very high degree of parallelism is still obtained. The execution in each iteration is deterministic simplifying the design of any control. In addition, since these algorithms rely heavily on matrix-vector multiplications, the structure of $\mathbf{A}$, being a binary matrix consisting of only 0/1 entries as illustrated in (\ref{eq:aexample}), can be readily utilized.

\subsubsection{Fast Projected Gradient for Solving NNLS -- Fast}
The fast projected gradient algorithm \cite{Polyak2015}, is described as
\begin{IEEEeqnarray}{rCl}
	\mathbf{x}^{(0)} & = &  \mathbf{0}, \ \mathbf{p}^{(0)} = \mathbf{0} \label{eq:initialization} \\
	\mathbf{x}^{(k+1)} & = & \max\left\{\mathbf{0},\ \mathbf{p}^{(k)}-\frac{1}{L}\mathbf{A}^T\left(\mathbf{A} \mathbf{p}^{(k)} -\mathbf{b}\right)\right\}  \label{eq:xupdate}\\
	\mathbf{p}^{(k+1)} & = & \mathbf{x}^{(k+1)} + s^{(k)}\left(\mathbf{x}^{(k+1)}- \mathbf{x}^{(k)}\right), \label{eq:pupdate}
\end{IEEEeqnarray}
where
\begin{equation}
c^{(k)} = \frac{1+\sqrt{1+4\left(c^{(k-1)}\right)^2}}{2}, \ s^{(k)} = \frac{c^{(k-1)}-1}{c^{(k)}},
\label{eq:sk}
\end{equation}
with $c^{(-1)} =  0$. $L$ is the Lipschitz constant which controls the step size in each iteration. It should be larger or equal to the spectral radius of $\mathbf{A}^T\mathbf{A}$, i.e., the largest absolute eigenvalue. For a given $\mathbf{A}$, this can be computed off-line. Replacing $\mathbf{p}^{(k)}$ with $\mathbf{x}^{(k)}$ in (\ref{eq:xupdate}) and not using (\ref{eq:pupdate}) and  (\ref{eq:sk})  gives the standard projected gradient algorithm for NNLS. The fast part comes from (\ref{eq:pupdate}), where the previous iteration value is used to accelerate the convergence. $s^{(k)}$ is the step length of the fast part and $\mathbf{p}^{(k)}$ is now denoted the predictor \cite{Polyak2015}.

Note here that the core of the algorithm, the matrix-vector multiplications, will be simple additions since the elements of $\mathbf{A}$ are 0 or 1 as seen from (\ref{eq:aexample}). As such, the architectures can not be directly used for NNLS problems without this property.

\subsubsection{Multiplicative Updates for Solving NNLS -- Mult}
In the multiplicative updates algorithm, the updates in each iteration are instead performed using multiplications and divisions and is described as \cite{Daube1986,Lee2001,Sha2007}:	
\begin{IEEEeqnarray}{rCl}
\mathbf{x}^{(0)} & = & \mathbf{1} \\
\mathbf{d} & = & \mathbf{A}^T\mathbf{b}\label{eq:mul2} \\
\mathbf{e}^{(k+1)} & = & \mathbf{A}^T\mathbf{A}\mathbf{x}^{(k)}\label{eq:mul3} \\
\mathbf{x}^{(k+1)} & = & \mathbf{d}\odot \mathbf{x}^{(k)} \oslash {\mathbf{e}^{(k+1)}}, \label{eq:mul5}
\end{IEEEeqnarray}
where $\mathbf{1}$ denotes a vector with all ones and $\odot$ and $\oslash$ represent component-wise (Hadamard) vector multiplication and division, respectively.

Similar to the Fast algorithm, the matrix-vector multiplications consist of additions.

\subsection{Proposed Architectures\label{sec:proposedarchitecture}}
The proposed architectures are iso-morphic mappings of a single iteration of the respective algorithms. Hence, each clock cycle one complete iteration of the algorithms is performed.
The primary reason to do this is to obtain a high degree of parallelism and to easily benefit from the fact that $\mathbf{A}$ only consists of 0 and 1 entries, leading to additions only.
Since the matrices $\mathbf{A}$ and $\mathbf{A}^T$ are constant in all iterations, these constant matrix-vector multiplications can be converted to summation structures where sub-expression sharing can be used to
reduce the area and power consumption of the implementation.


\subsubsection{Fast }
\begin{figure}
	\centering
	\includegraphics[scale=1.2]{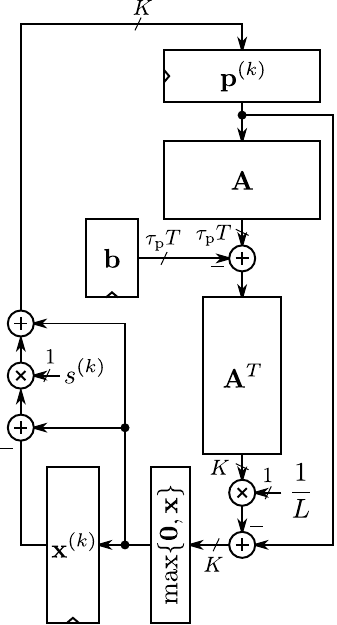}
	\caption{Proposed architecture of Fast algorithm with number of parallel signals annotated.}
	\label{fig:architecture}
\end{figure}

The resulting architecture is shown in Fig.~\ref{fig:architecture}.
It should be noted that most signals are vectors.
There are registers to store $\mathbf{b}$, which is given at the start of a detection process, and $\mathbf{p}^{(k)}$ and $\mathbf{x}^{(k)}$, which are initialized to zero at the start as seen in (\ref{eq:initialization}).

To implement multiplication by $s^{(k)}$, the value of it should be calculated in each iteration and then multiplied. As visible from (\ref{eq:sk}), the calculation of $s^{(k)}$ is costly in hardware, but it is possible to compute the $s^{(k)}$ values off-line and save them in a memory. Furthermore, $s^{(k)}$ converges to one after some time, so for $k$ and all later iterations, the values will saturate to the largest representable number less than or equal to one.

\subsubsection{Mult}
To avoid the relatively costly multiplications and divisions of the Mult algorithm, a logarithmic number system (LNS) \cite{Swartzlander1975} is used for those parts. This reduces the multiplications and divisions to additions and subtractions, respectively, at the expense of converting the numbers between the linear and logarithmic domains.

The multiplicative updates algorithm, partly in the logarithmic domain, is described as:
\begin{IEEEeqnarray}{rCl}
\hat{\mathbf{x}}^{(0)} & = & \mathbf{0} \\
\hat{\mathbf{d}} & = & \log \left(\mathbf{A}^T\mathbf{b}\right) \\
\hat{\mathbf{e}}^{(k+1)} & = & \log \left(\mathbf{A}^T\mathbf{A}\mathbf{x}^{(k)} \right) \\
\hat{\mathbf{x}}^{(k+1)} & = & \hat{\mathbf{d}} + \hat{\mathbf{x}}^{(k)} - {\hat{\mathbf{e}}^{(k+1)}} \\
\mathbf{x}^{(k+1)} & = & \exp \left(\hat{\mathbf{x}}^{(k+1)}\right).
\end{IEEEeqnarray}

\begin{figure}
	\centering
	\includegraphics[scale=1.2]{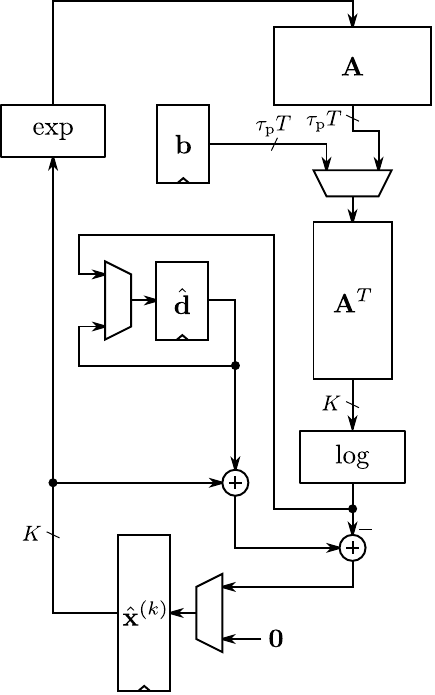}
	\caption{Proposed architecture of Mult algorithm partially implemented in the logarithmic domain with number of parallel signals annotated.}
	\label{fig:MULarchitecture}
\end{figure}

The resulting architecture is shown in Fig.~\ref{fig:MULarchitecture}. Note that $\hat{\mathbf{d}}$ is computed in the first clock cycle and that the first iteration is performed in the second clock cycle. The multiplexers are used for storing the computed value of $\hat{\mathbf{d}}$ and to initialize $\hat{\mathbf{x}}^{(0)}$ in the first clock cycle. Hence, one more clock cycle is required here compared to the Fast algorithm for the same number of iterations.

To compute the logarithm and exponentiation, either a table or using Mitchell's logarithm approximation \cite{Michell1962} is considered. There are many methods suggested to either perform approximations of the logarithm and exponent functions using smaller tables or to improve the accuracy of Mitchell's algorithm. However, as seen in the results section, the accuracy of the Mitchell approximation is enough.

The idea of Mitchell's logarithm approximation is to utilize the approximation
\begin{equation}\label{eq:logapprox}
\log (x) \approx x - 1
\end{equation}
in the range $1 \leq x \leq 2$.

To benefit from this, the position of the leading one, i.e., the most significant bit being one, of the input $x$ is determined\footnote{Here, the input is always non-negative, so the negative case where the leading zero should be detected does not have to be considered.} and then shifted to the range $1 \leq x < 2$. The position of the first one also determines the integer part of the logarithm. Then, the leading one is removed and the fractional part of the logarithm is obtained as per (\ref{eq:logapprox}).

For computing the exponent, the opposite procedure is performed. A one is put in front of the fractional part and the final result is obtained by shifting the fractional result with the leading one according to the number of integer bits.


\subsubsection{A Note on Pipelining}
It should be noted here that pipelining will not provide any advantage as we need the result of one iteration before being able to start the next one.
If required, one may interleave several detections by introducing pipelining, although that is not considered here since it will increase the detection latency and storage requirements.
Unfolding may provide a small increase of detection rate as unfolding $N$ times will most likely not decrease the clock frequency $N$ times since the synthesis tool probably can optimize some parts better.
Naturally, fully unfolding the architecture the required number of iterations will enable pipelining and streaming processing.
However, the obtained detection rate is most likely more than enough for most practical current and envisioned scenarios.

\section{Results}\label{sec:mainresults}
For the implementation example, we consider a system with parameters as shown in Table~\ref{tab:param}. These parameters are selected based on the number of users and user activity, where the number of antennas, orthogonal pilots, and transmitted pilots are found by simulation to obtain a reasonable detection performance. For more results on how these parameters relate, we refer to \cite{Becirovic2019}. Synthesis results are for 28~nm FD-SOI standard cells with a power supply voltage of 1.0~V, using Design Compiler. Power estimates for the final designs are obtained by simulating the synthesized netlist using SDF data to produce switching activities that are back annotated into the synthesis tool. Compared to the preliminary version \cite{Sarband2020}, here the low leakage library with a slow-slow characterization at 125$^\circ$C is used.

\begin{table}
	\centering
	\caption{Parameters for the Implemented Example.\label{tab:param}}
	\begin{tabular}{l|r}
		Parameter & Value \\
		\hline
		Number of users, $K$ & 1024 \\
		User activity & $2^{-6}$\\
		Number of antennas, $M$ & 96 \\
		Number of orthogonal pilots, $\tau_\text{p}$ & 16 \\
		Number of transmitted pilots, $T$ & 8 \\
	\end{tabular}
\end{table}

\subsection{Number of Iterations}
Determining the number of iterations required for robust activity detection is important since this relates the clock frequency of the implementation with the number of detections per second. Here, we will consider floating-point implementation of the algorithms, while fixed-point implementation is further discussed in Section~\ref{sec:wordlength}.

\begin{figure}
	\centering
	\includegraphics[scale=.3]{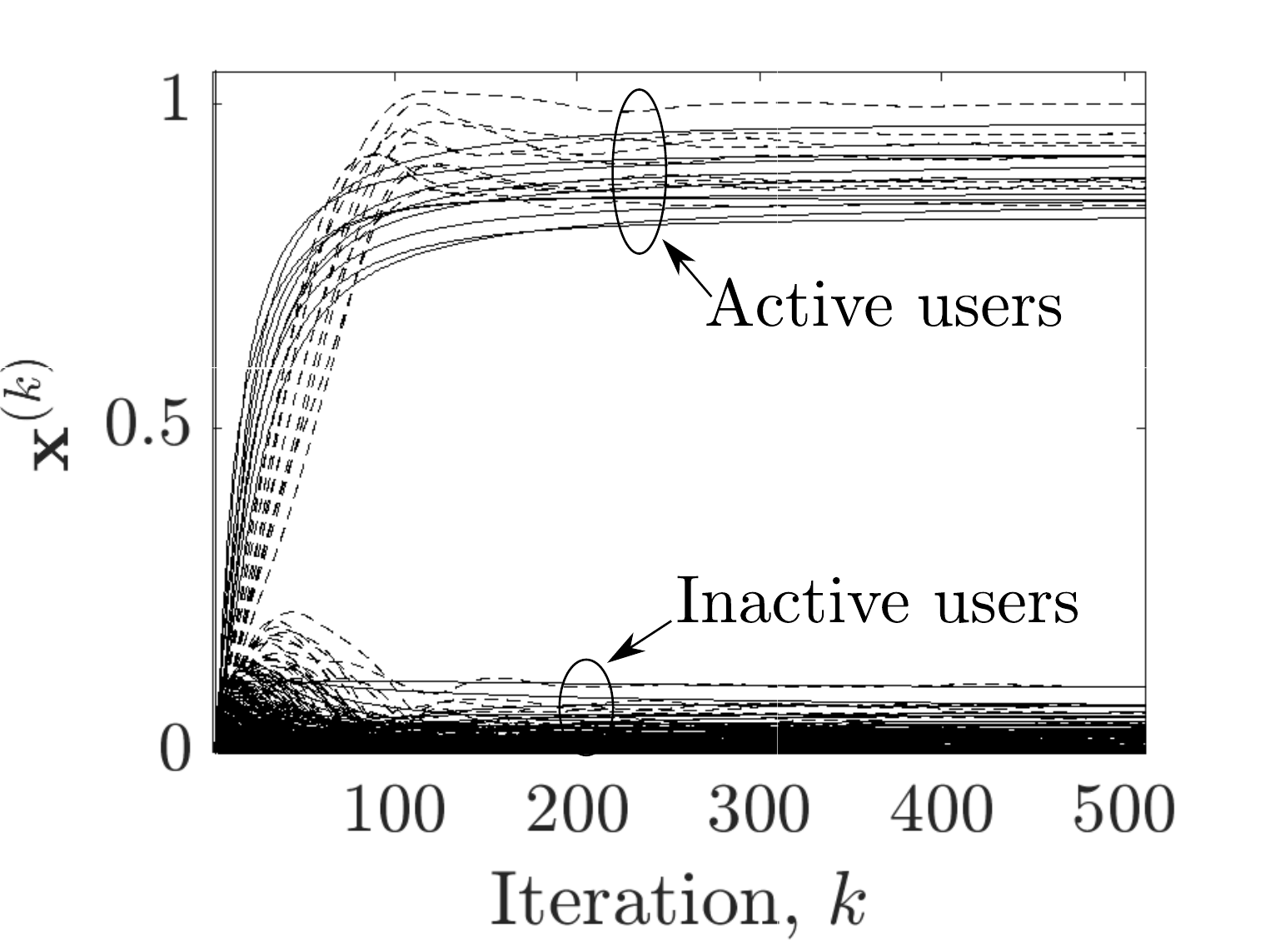}
	\caption{Convergence of $\hat{\mathbf{x}}$ using  Fast (dashed) and Mult (solid) algorithms successfully performing an activity detection. Large values corresponds to active users and small values to inactive users. There is a clear separation between the active and inactive users.}
	\label{fig:compconverge}
\end{figure}

First, the behavior of the algorithms is illustrated. When iterating the algorithms, the values in $\hat{\mathbf{x}}$ corresponding to active users will be close to one, while the values corresponding to inactive users will be close to zero. A typical sequence of values for a successful decoding is shown in  Fig.~\ref{fig:compconverge}, where the values of  $\hat{\mathbf{x}}$ are shown after each iteration. Here, one can see that the Mult algorithm initially has a faster convergence for active users, but that around iteration 50, the Fast algorithm has caught up and reaches the final values slightly faster. One can see that with about 100 iterations, it is possible to clearly separate the active and inactive users.

\begin{figure}
	\centering
	\includegraphics[scale=.3]{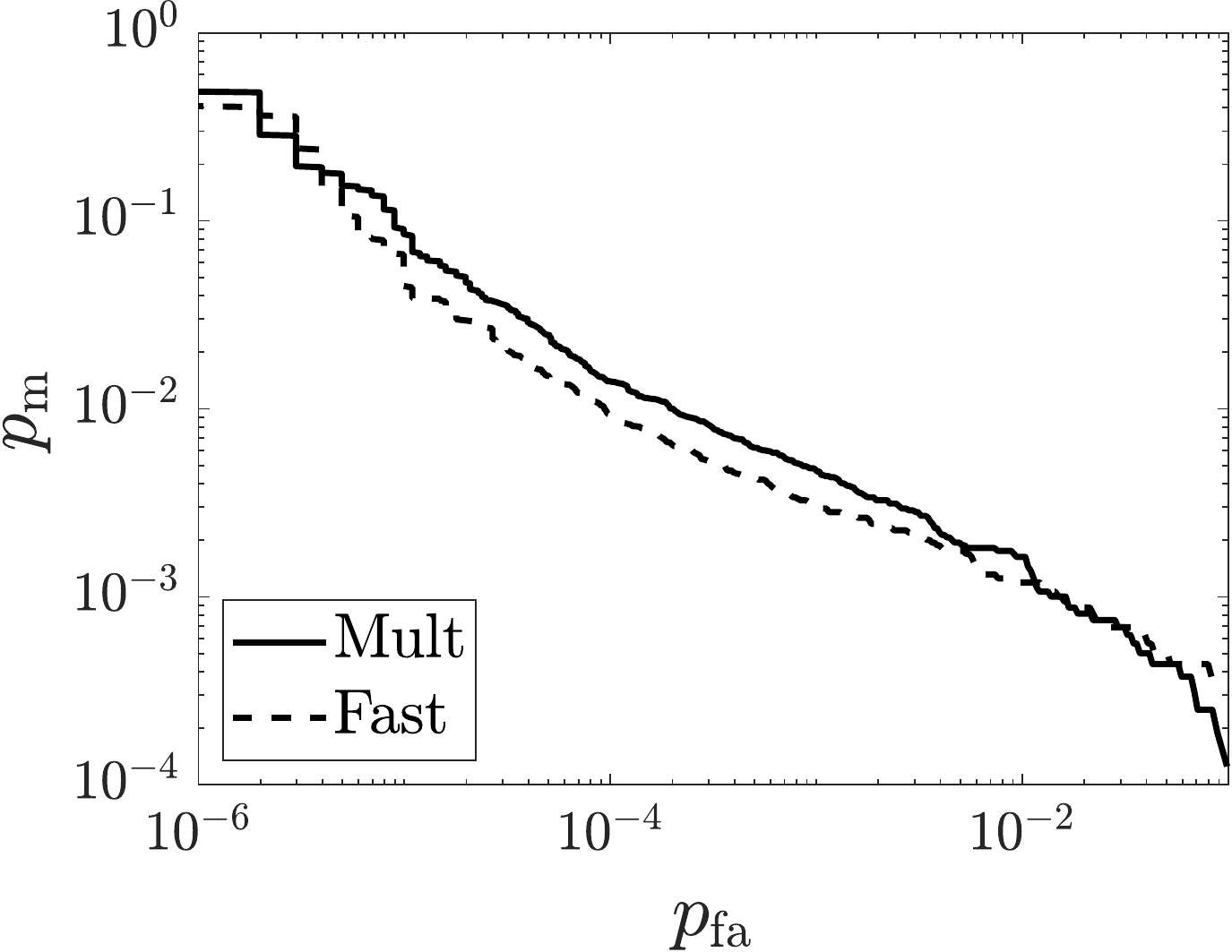}
	\caption{Missed detection, $p_{\text{m}}$, versus false alarm rates, $p_{\text{fa}}$, for 4000 iterations results of floating-point realizations of Fast and Mult algorithms.}
	\label{fig:ROCconvergedfloating}
\end{figure}

Typically, one does not use a fixed-threshold value, but studies the effect of the number of active users that are not detected (missed detections) and number of inactive users that are detected as active (false alarms) with a varying threshold.
The probability of missed detections, $p_{\text{m}}$, and false alarm, $p_{\text{fa}}$, are calculated as
\begin{IEEEeqnarray}{rCl}
	p_{\text{m}}& =\frac{\#\text{undetected active users}}{\#\text{active users}}    \label{eq:pm}\\
	p_{\text{fa}}& =\frac {\#\text{detected inactive users}} {\#\text{inactive users}}\label{eq:pfa}
\end{IEEEeqnarray}
and plotted against each other in a receiver operating characteristic (ROC) curve. The ROC curves for the two considered algorithms are shown in Fig.~\ref{fig:ROCconvergedfloating} when using 4000 iterations, clearly more than in Fig.~\ref{fig:compconverge}, and a large number of instances. Here, it can be seen that the Fast algorithm has a slightly better behavior compared to the Mult algorithm. Each point on the curve corresponds to a threshold and the position, the number of missed detections and false alarms if that threshold was used. Hence, the values in the top left corner corresponds to large thresholds and the values in the bottom right corner corresponds to small thresholds. For example, if a threshold of, say, 0.9 is used, there will almost never be any false alarms, but it is quite common that there are missed detections.

From Fig.~\ref{fig:compconverge}, one can realize that for good convergence cases, it is possible to get a good separation of the active and inactive users is a few hundred iterations.
 
\begin{figure}
	\centering
	\subfigure[]{\label{fig:badfast}\includegraphics[scale=.28]{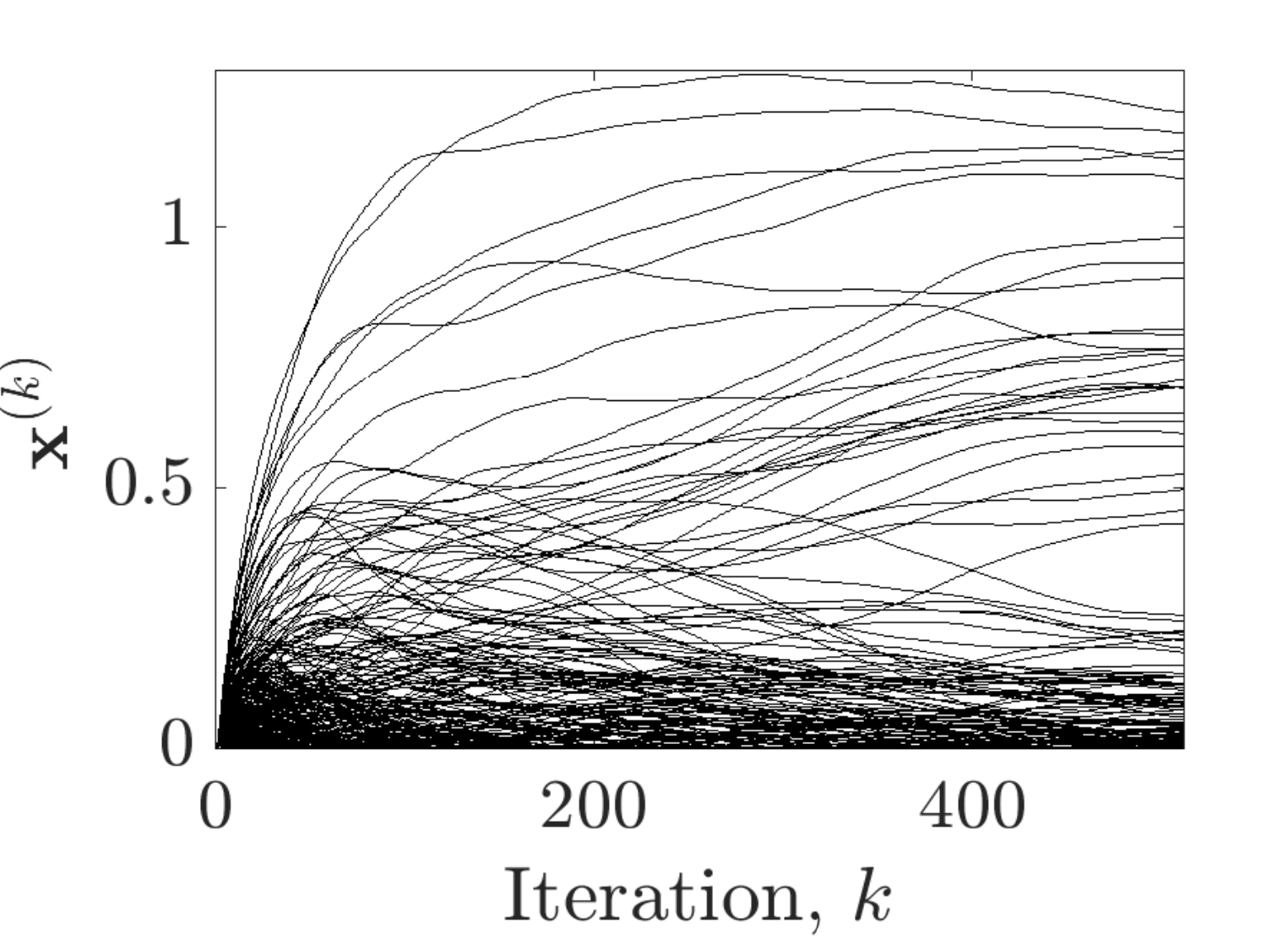}} 
	\hspace{-2mm} 
	\subfigure[]{\label{fig:badmul}\includegraphics[scale=.28]{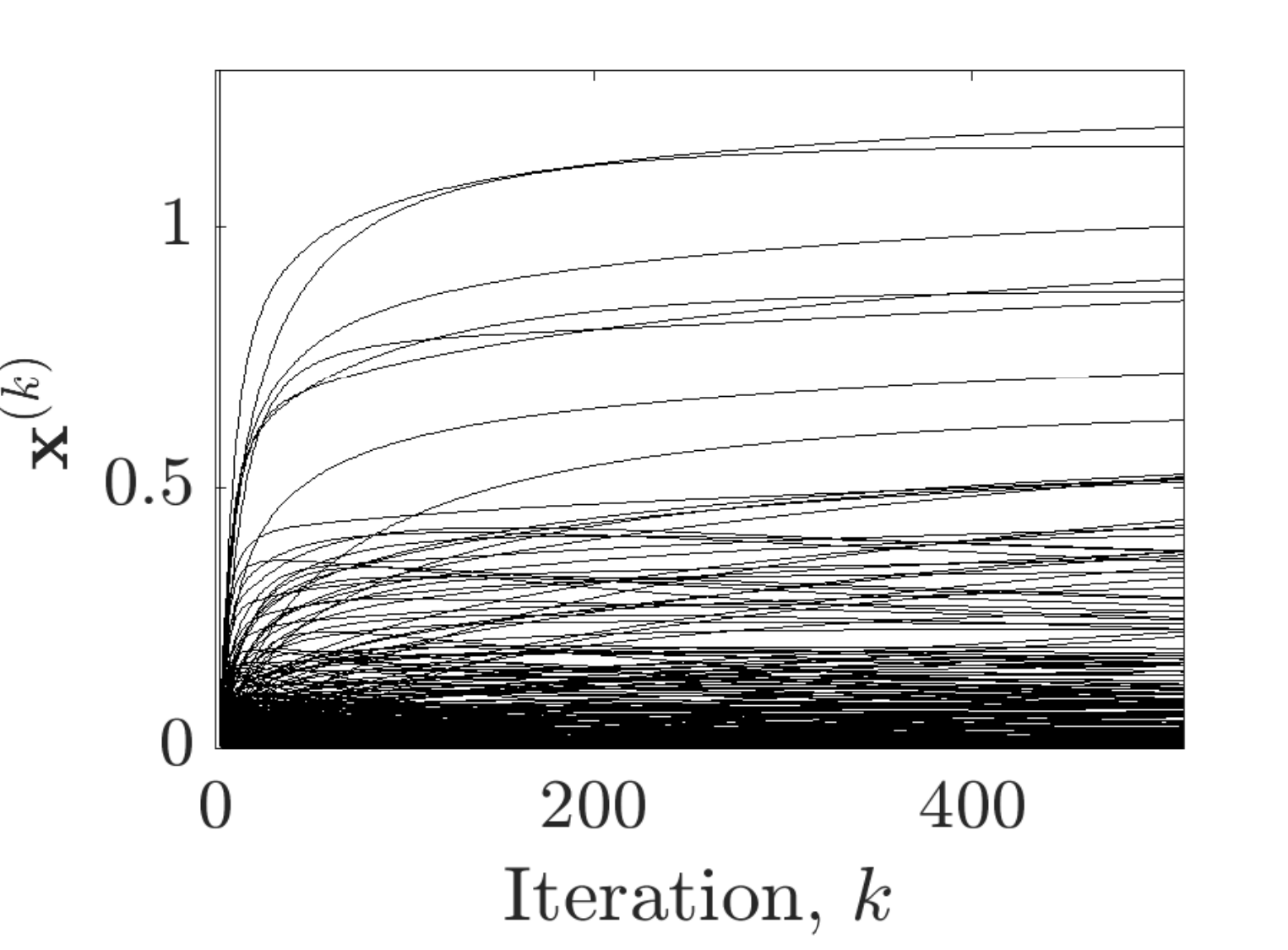}}   
	\caption{Convergence of $\hat{\textbf{x}}$ using  (a) Fast and (b) Mult algorithms where the active and inactive users are not clearly separated. 
	}
	\label{fig:badconverge}
\end{figure}

Instead, consider a case with bad convergence, as shown in Fig.~\ref{fig:badconverge}. Here, the channel conditions and the selection of active users interacts in such a way that it is very hard to distinguish the active and inactive users. In these cases, one can expect that the curves of some active users are actually below the curves of some inactive users and that there are no distinct value that will separate the two groups.

\begin{figure}
	\centering
	\subfigure[]{\label{fig:badfast2}\includegraphics[scale=.28]{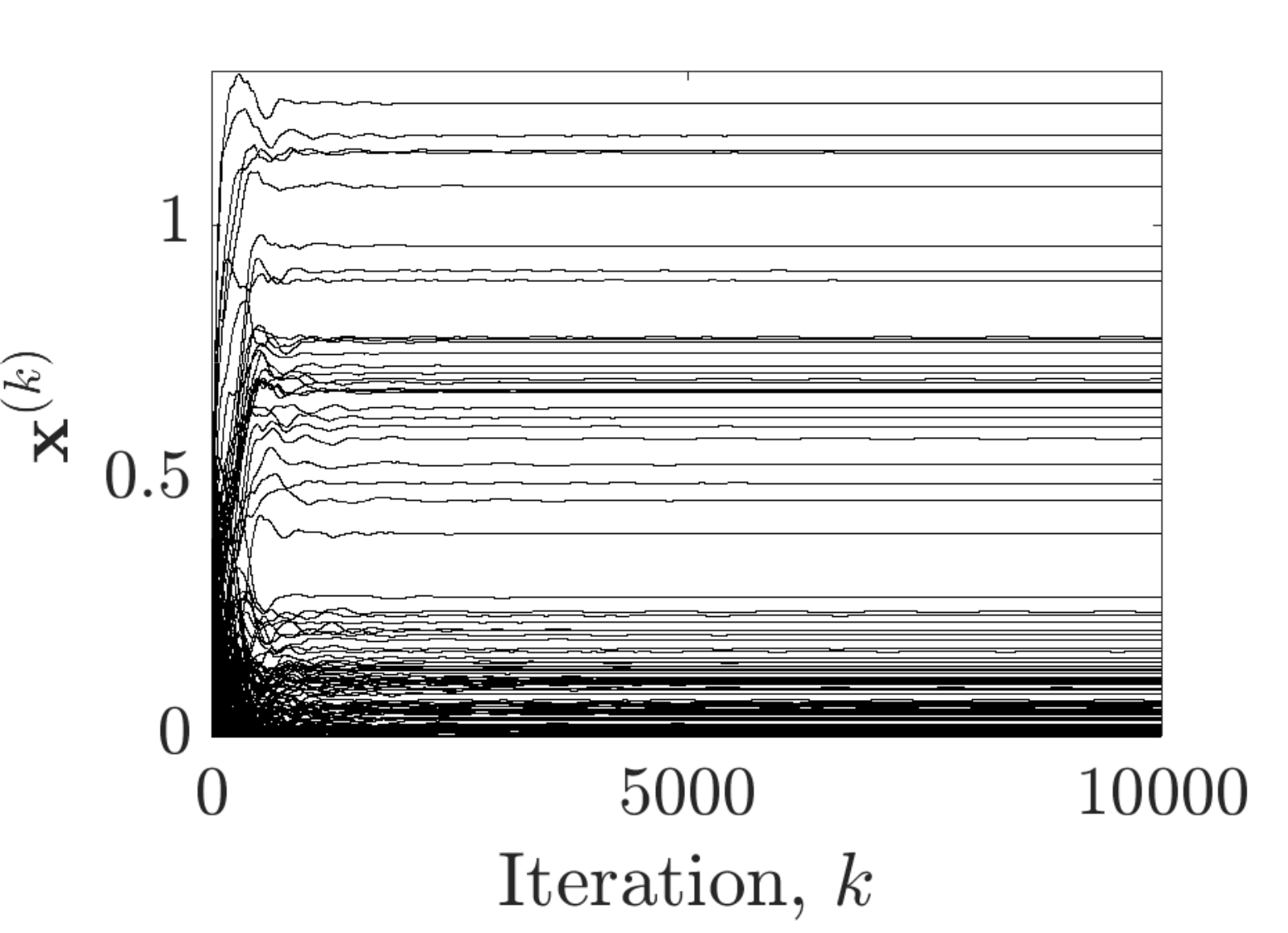}}\hspace{-2mm} 
	\subfigure[]{\label{fig:badmul2}\includegraphics[scale=.28]{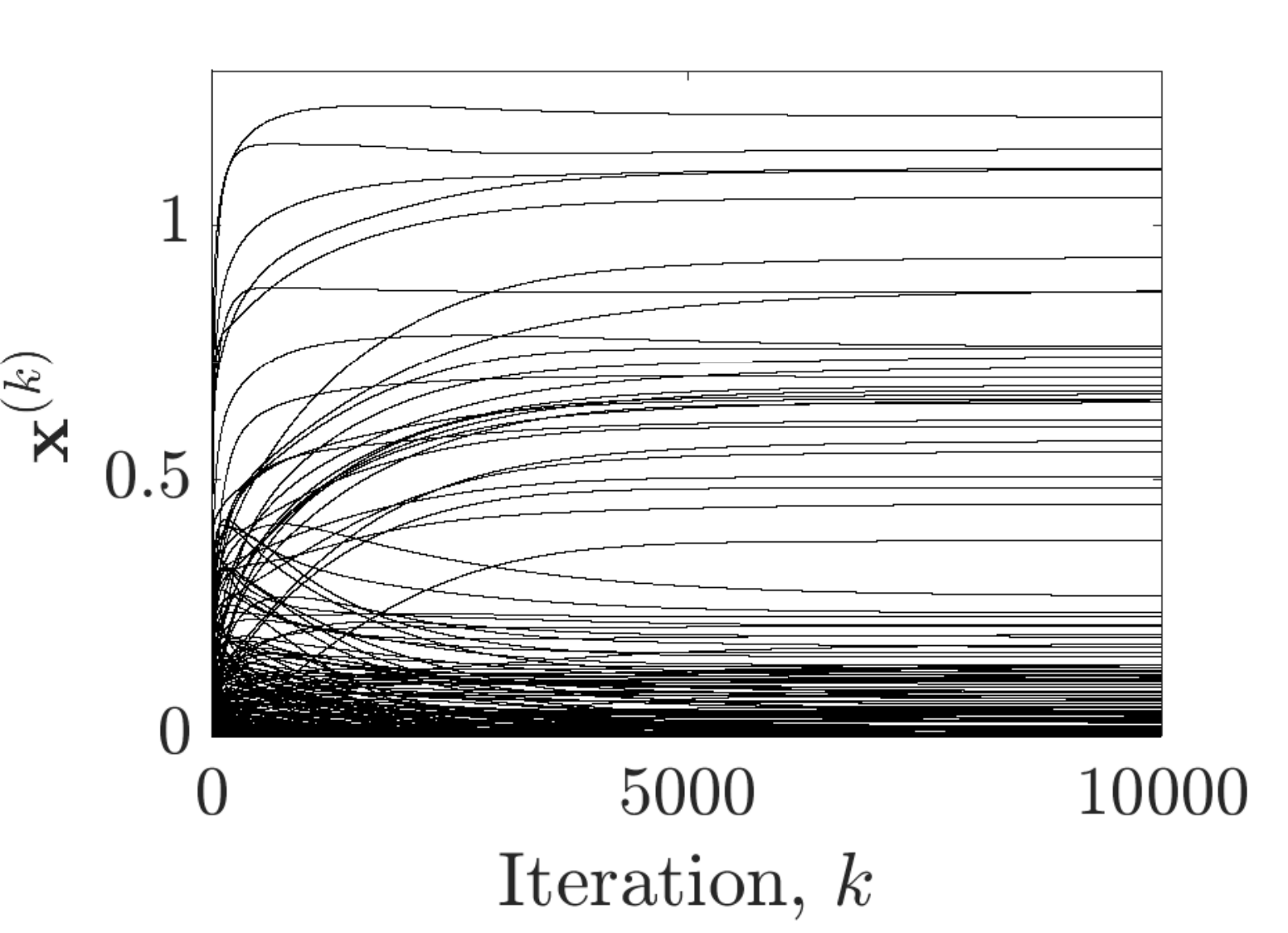}}   
	\caption{Convergence of $\hat{\textbf{x}}$ using  (a) Fast and (b) Mult algorithms where the active and inactive users are not clearly separated. Extended version of Fig.~\ref{fig:badconverge} covering 10000 iterations.
	}
	\label{fig:badconverge2}
\end{figure}

To study this further, the same case as in Fig.~\ref{fig:badconverge}, was solved using 10000 iterations, with the results shown in Fig.~\ref{fig:badconverge2}. Here, it can be seen that the Fast algorithm in Fig.~\ref{fig:badfast2} reaches a more or less stable value in about 1000 iterations, while for the Mult algorithm in Fig.~\ref{fig:badmul2} some variables have not reached a stable value even after 10000 iterations.

Therefore, it can be argued that although Mult gives worse results in Fig.~\ref{fig:ROCconvergedfloating}, the degradation comes primarily from cases where the results will contain detection errors anyway. Using even more iterations will make the curves in Fig.~\ref{fig:ROCconvergedfloating} identical. Hence, we will use the Fast algorithm curve in Fig.~\ref{fig:ROCconvergedfloating} as a reference curve for a converged solution.

\begin{figure}
	\centering
	\subfigure[]{\label{fig:ROCfastfloatIT}\includegraphics[scale=0.3]{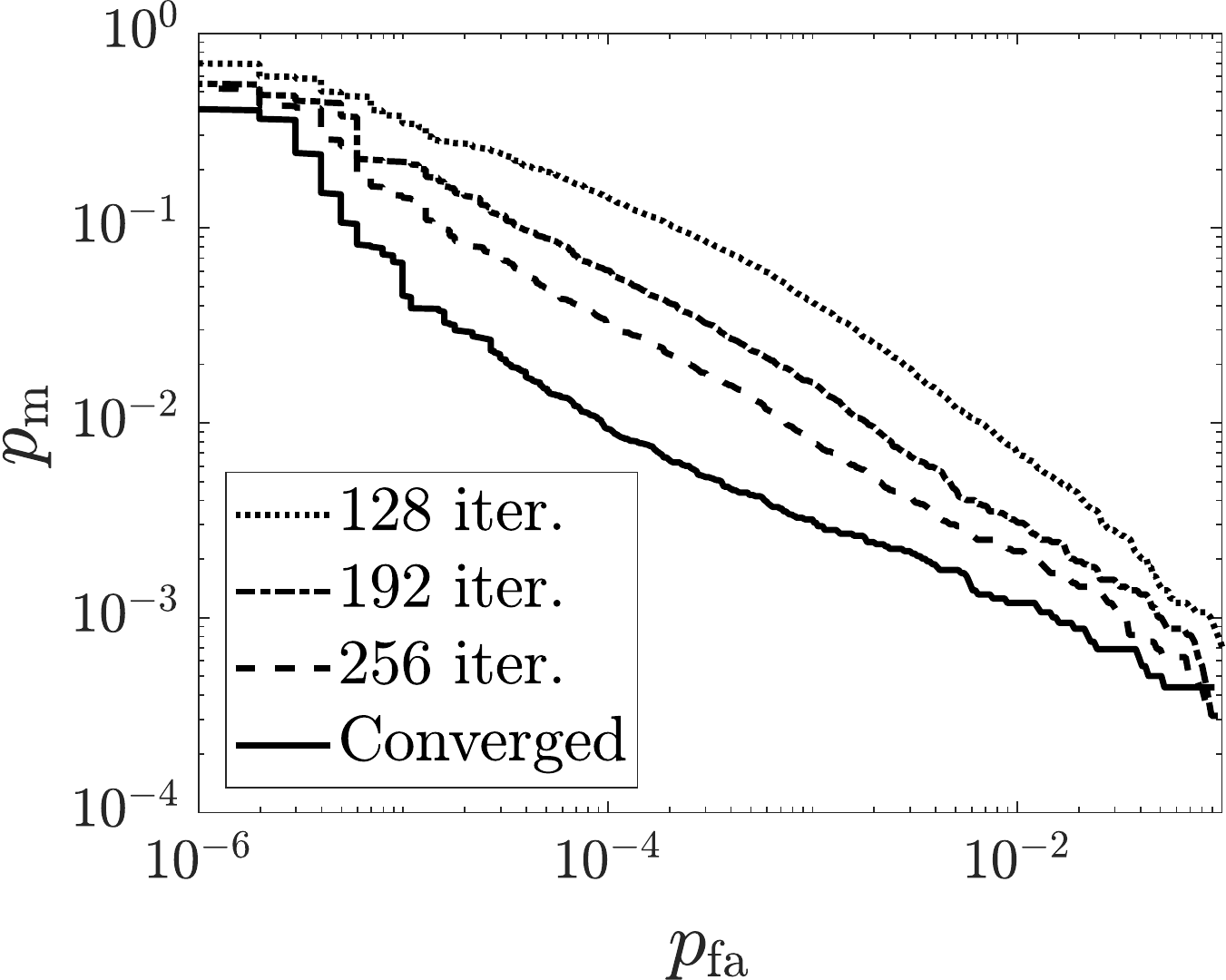}}
	\hspace{-1mm} 
	\subfigure[]{\label{fig:ROCMULfloatIT}\includegraphics[scale=.3]{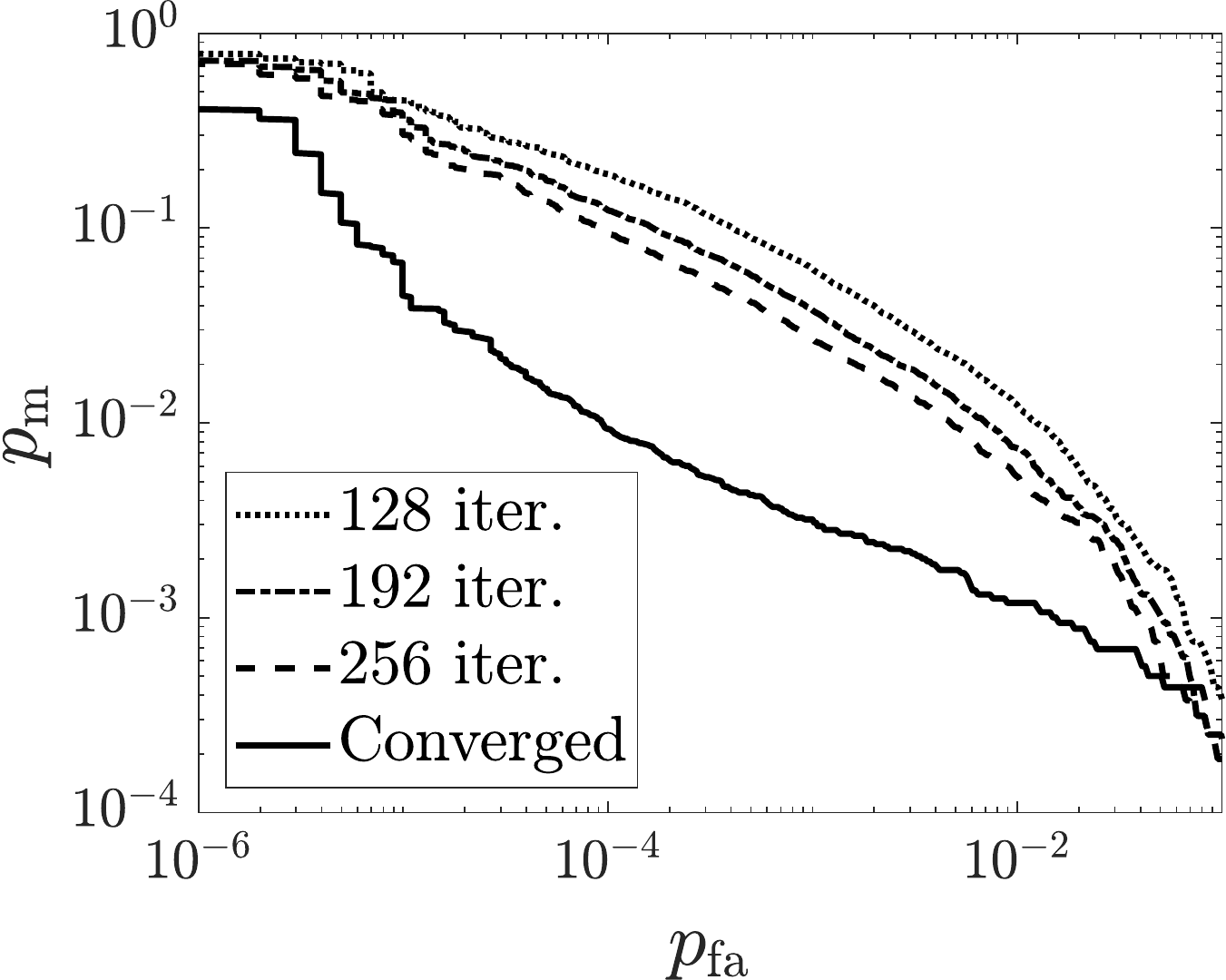}}
	\caption{Missed detection, $p_{\text{m}}$, versus false alarm rates, $p_{\text{fa}}$, for floating-point implementations, 128 iterations, 192 iterations and 256 iterations for (a) Fast algorithm, and (b) Mult algorithm.}
	\label{fig:ROCiteretedfloating}
\end{figure}

In Fig.~\ref{fig:ROCiteretedfloating}, the missed detection, $p_{\text{m}}$, and false alarm rates, $p_{\text{fa}}$, are shown for the floating-point implementations with 128, 192, and 256 iterations,  for the Fast and Mult algorithms. As seen, the results are better for the Fast algorithm, but we know from previous discussions that the Mult algorithm converges faster in the good cases, so this will also be further considered.

%

\subsection{Word Length Optimization}\label{sec:wordlength}
A careful word length optimization has been performed to determine the minimum resolution of the signals while still being able to operate correctly.
In addition, the maximum values have been determined by a combination of interval arithmetic and simulations.
One example of the latter is the output of the $\mathbf{A}$ block that theoretically may require seven additional MSBs compared to the input as up to 87 $\mathbf{p}^{(k)}$\mbox{-}values are summed for the selected $\mathbf{A}$ since the maximum number of ones in a row of the selected $\mathbf{A}$ is 87.
However, simulation show that at most five additional MSBs are required in practice as not all the $\mathbf{p}^{(k)}$\mbox{-}values will be close to one.
Similarly, for the Fast algorithm, the subtraction of $\mathbf{x}^{(k)}$ from $\mathbf{x}^{(k+1)}$ results in a shorter output word length than the input word length. It can be seen from Figs.~\ref{fig:compconverge} and \ref{fig:badconverge} that the values are correlated and do not change much between two iterations.

In the following, $S$ and $U$ denote signed and unsigned representation, respectively.
The numbers following $S$ and $U$ are the number of integer and fractional bits, respectively.
A negative number of integer bits denotes that the corresponding number of fractional bits are not required. As examples, $S(2,2)$ denotes a four-bit signed number with values between $ -\frac{2^{4-1}}{2^2} = -2$ and $\frac{2^{4-1}-1}{2^2} = 1.75$ and $U(-2, 4)$ denotes a two-bit unsigned number with values between $0$ and $\frac{2^2-1}{2^4} = 0.1875$. In general, a $S(I,F)$ number goes from $-\frac{2^{I+F-1}}{2^F} = -{2^{I-1}}$ to $\frac{2^{I+F-1}-1}{2^F} = 2^{I-1} - {2^{-F}}$ and a $U(I,F)$ number goes from $0$ to $\frac{2^{I+F}-1}{2^F} = 2^{I} - {2^{-F}}$.

For the nodes where quantization is performed, i.e., the number of fractional bits are reduced, rounding is used.

\begin{figure}
	\centering
	\subfigure[]{\label{fig:ROCfast}\includegraphics[scale=0.3]{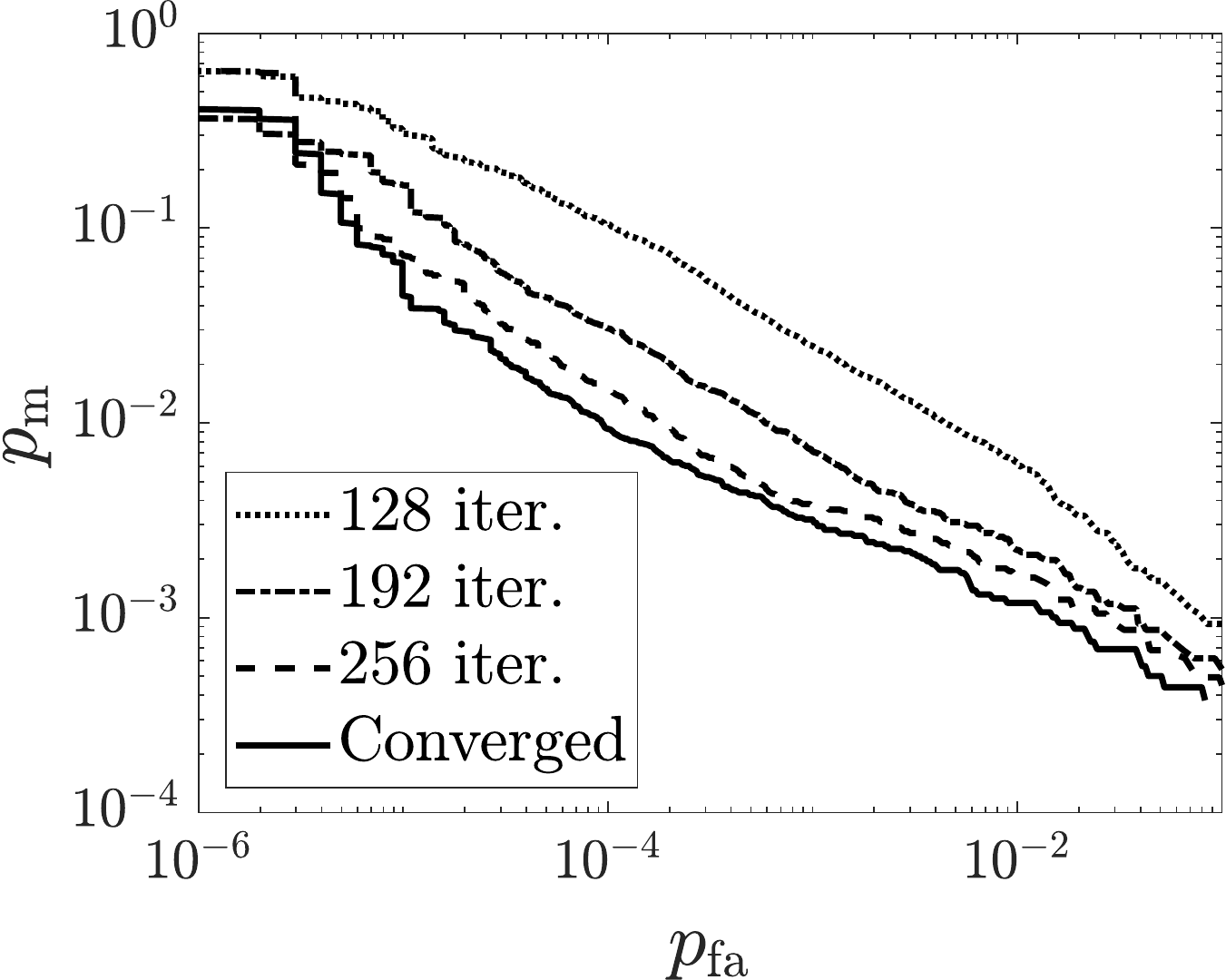}}
	\vfill
	\subfigure[]{\label{fig:ROCMULtable}\includegraphics[scale=0.3]{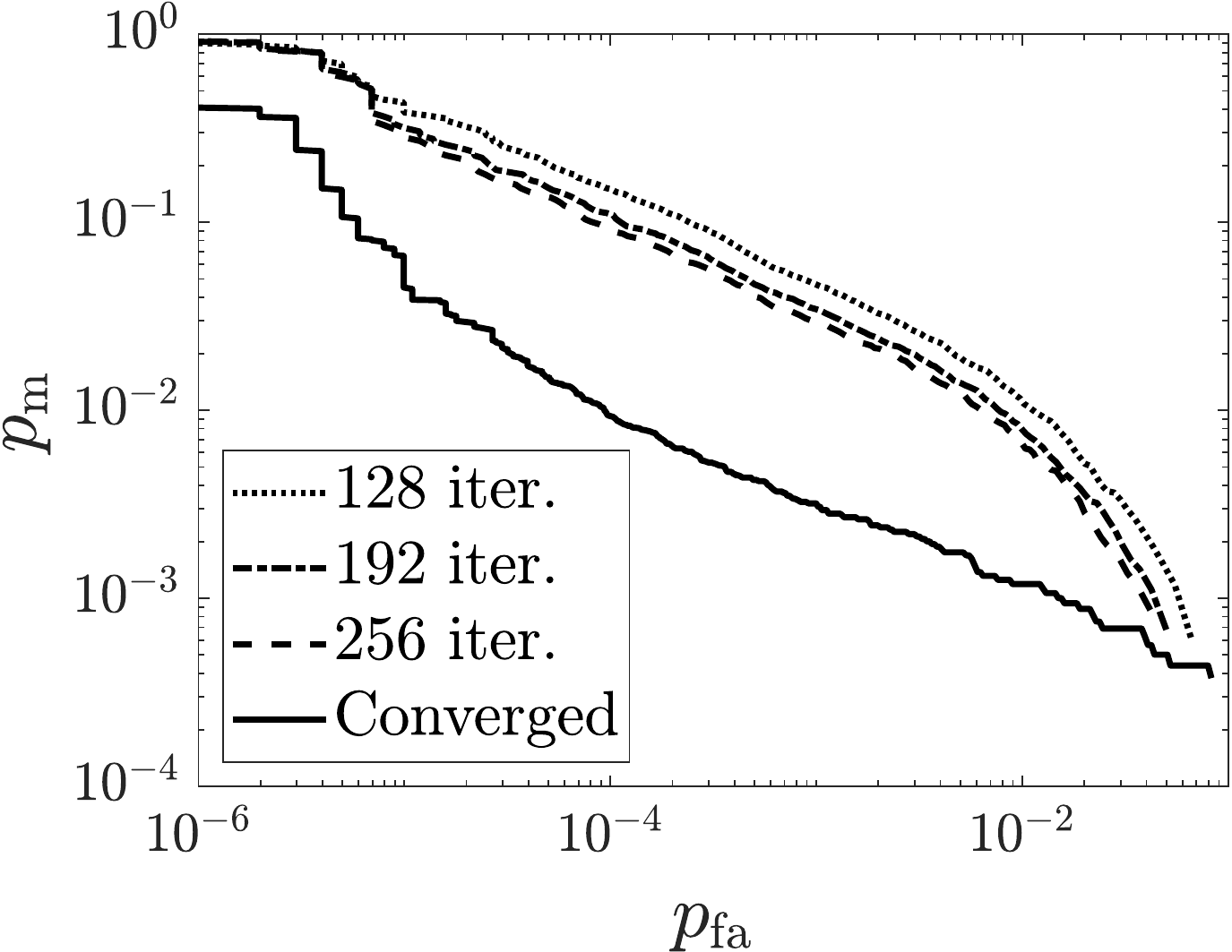}}
	\hspace{-1mm} 
	\subfigure[]{\label{fig:ROCMULMitchell}\includegraphics[scale=.3]{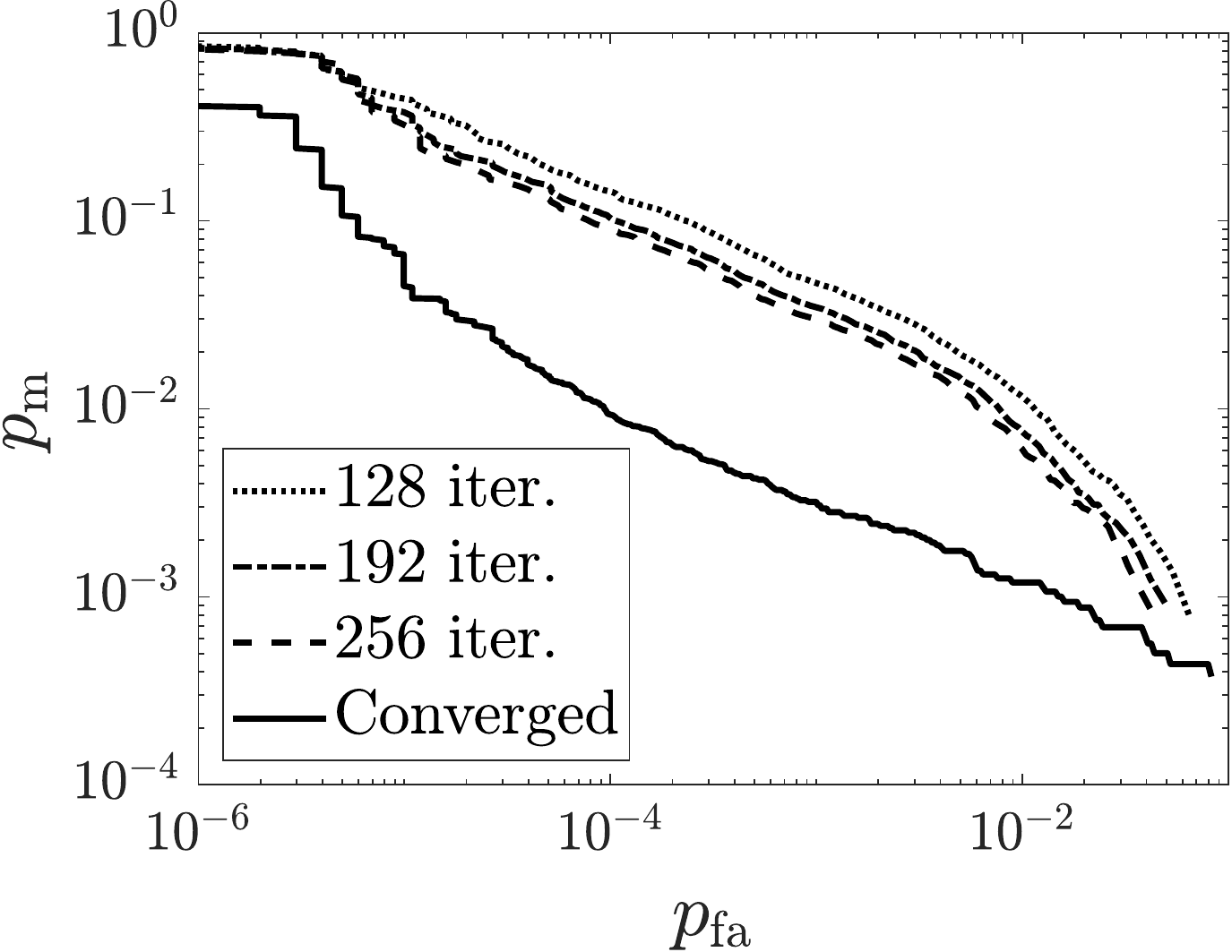}}
	
	\caption{Missed detection, $p_{\text{m}}$, versus false alarm $p_{\text{fa}}$, rates for converged floating-point and fixed-point with 128, 192, and 256 iterations, for (a) Fast algorithm, (b) Mult algorithm using lookup tables for logarithm and exponent computations, and (c) Mult algorithm using Mitchell's approximated logarithm.}
	\label{fig:ROCfloatfixed} 
\end{figure}

The resulting ROC curves for the resulting word lengths, further discussed below, are shown in Fig.~\ref{fig:ROCfloatfixed}. Here, one can see that 256 and 192 iterations are required for the Fast and Mult algorithms to reach close to the floating-point converged performance, respectively. On the other hand, 128 iterations are enough for the Fast algorithm to reach a performance similar to the Mult algorithm.

\subsubsection{Fast}
\begin{figure}
	\centering
	\includegraphics[scale=1.2]{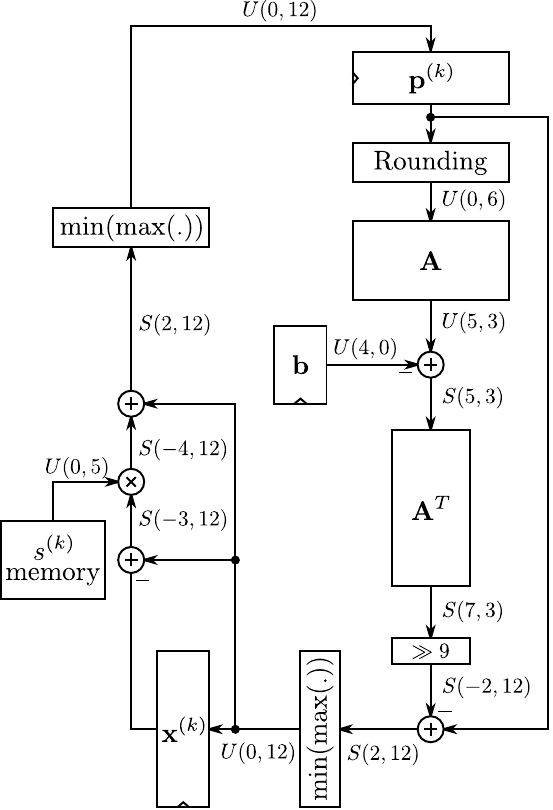}
	\caption{Modified proposed architecture for the Fast algorithm with annotated word lengths for the implemented example.}
	\label{fig:wl}
\end{figure}

The value of $L$ for the considered $\mathbf{A}$ is 520.74.
Strictly, this should be rounded up to guarantee convergence under all possible inputs, but simulations show that selecting $L = 512$ works well, resulting in a shift instead of a multiplication.
The values of $s^{(k)}$ are pre-computed and stored in a lookup table (all values for $k \geq 58$ are $1-2^{-5}$).
The values for $\mathbf{p}^{k}$ closely follow the values of $\mathbf{x}^{k}$.
As seen from Fig.~\ref{fig:badfast}, sometimes the values for $\mathbf{x}^{k}$, and therefore $\mathbf{p}^{k}$, can take on values above one.
Hence, we saturate the values to at most one, using a single guard-bit. This operation is combined with the $\max$ operation in (\ref{eq:xupdate}) to the $\min(\max(.))$-operation in Fig.~\ref{fig:wl}. 
It can also be noted that (\ref{eq:pupdate}) on rare occasions result in a small negative value, which is confirmed by simulations.
Hence, $\min(\max(.))$-operations are performed before storing $\mathbf{p}^{k}$ as well to keep the word length limited and enabling storage of unsigned values only.

The resulting architecture with annotated word lengths is shown in Fig.~\ref{fig:wl}.
Rounding is here also performed for the input to the matrix-vector multiplication by $\mathbf{A}$, which significantly reduces the complexity of the computation by halving the input word length. 

The implemented Fast algorithm can therefore be described as
\begin{IEEEeqnarray}{rCl}
	\mathbf{x}^{(0)} & = &  \mathbf{0}, \ \mathbf{p}^{(0)} = \mathbf{0} \nonumber \\
	\mathbf{x}^{(k+1)} & = & \min\left\{\mathbf{m},  \max\left\{\mathbf{0},\ \mathbf{p}^{(k)}-\frac{1}{L}\mathbf{A}^T\left(\mathbf{A} \mathbf{p}^{(k)} -\mathbf{b}\right)\right\} \right\} \nonumber\\
	\mathbf{p}^{(k+1)} & = &  \min\left\{\mathbf{m},\right.  \nonumber \\
& & 	\left.\max\left\{\mathbf{0},\ \mathbf{x}^{(k+1)} + s^{(k)}\left(\mathbf{x}^{(k+1)}- \mathbf{x}^{(k)}\right)\right\} \right\} \nonumber
\end{IEEEeqnarray}
where $\mathbf{m}$ is a vector of the largest positive value representable, $1-2^{-12}$.
\subsubsection{Mult}
As discussed earlier, the Mult architecture is partially based on computation in the logarithmic domain. The results in Figs.~\ref{fig:ROCMULtable} and \ref{fig:ROCMULMitchell} show that the Mitchell logarithm approximation gives the same activity detection performance as using a lookup table.

As part of the word length optimization, the initial value of $\mathbf{x}^{(0)}$ was reduced from 1 to $2^{-7}$. This reduces the values of $\mathbf{e}^{(1)}$, allowing a reduced word length, without reducing the detection performance. In addition, saturation was introduced in order to handle potential overflows of $\hat{\mathbf{x}}^{(k+1)}$.

\begin{figure}
	\centering
	\includegraphics[scale=1.2]{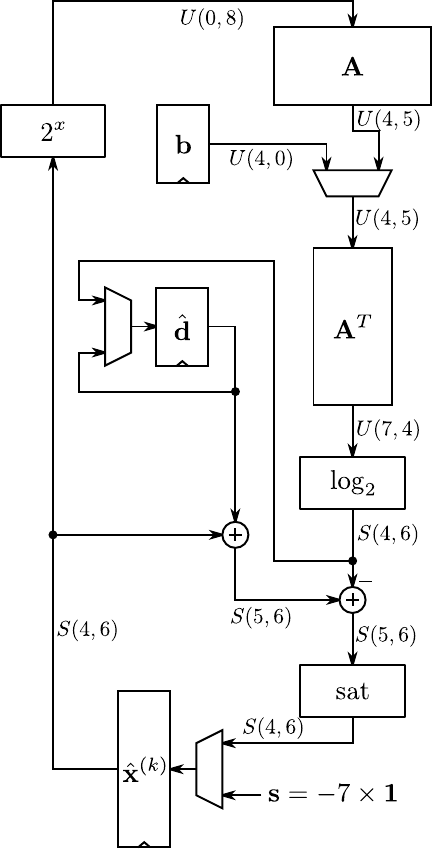}
	\caption{Modified proposed architecture for the Mult algorithm partially implemented in the logarithmic domain with annotated word lengths for the implemented example.}
	\label{fig:wlmul}
\end{figure}

This leads to that the implemented Mult algorithm, partially in the logarithmic domain, is described as
\begin{IEEEeqnarray}{rCl}
	\hat{\mathbf{x}}^{(0)} & = & \mathbf{s} \\
	\hat{\mathbf{d}} & = & \log_2 \left(\mathbf{A}^T\mathbf{b}\right) \\
	\hat{\mathbf{e}}^{(k+1)} & = & \log_2 \left(\mathbf{A}^T\mathbf{A}\mathbf{x}^{(k)} \right) \\
	\hat{\mathbf{x}}^{(k+1)} & = & \text{sat}\left(\hat{\mathbf{d}} + \hat{\mathbf{x}}^{(k)}_i - {\hat{\mathbf{e}}^{(k+1)}} \right) \\
	\mathbf{x}^{(k+1)} & = & 2^{\hat{\mathbf{x}}^{(k+1)}}
\end{IEEEeqnarray}
where $\mathbf{s}$ is a vector with the initial value, $\log_2\left(2^{-7}\right) = -7$, and $\mathrm{sat}$ corresponds to saturating the result in case of overflow. The resulting architecture with annotated word lengths is shown in Fig.~\ref{fig:wlmul}.

\begin{figure}
	\centering
	\includegraphics[scale=.7]{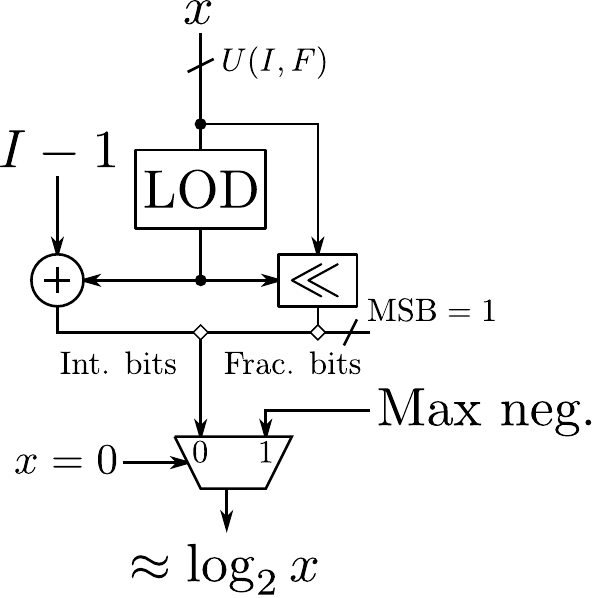} 
	\caption{Realization of the Mitchell logarithm approximation with a $U(I,F)$ input format and zero input saturating to the most negative output value.}
	\label{fig:mitchelllog2}
\end{figure}

The structure of the used logarithm approximation with base 2 is shown in Fig.~\ref{fig:mitchelllog2}. It consists of a leading one detector (LOD) that outputs the position of the first one in the input word, with the MSB corresponding to 0. As there in general are $I$ integer bits, the value must be offset by $I-1$ to determine the correct integer part of the logarithmic value. The LOD value is also used to left-shift the input such that the leading one comes in the MSB position. This leading one is removed and the remaining part is used as the fractional part of the logarithm and combined with the integer part. Finally, as a zero input value will output the most negative output value, a simple saturation multiplexer is added. The programmable shift is realized using a 11-to-1-multiplexer, which provided the best synthesis results.

The logarithm approximation in Fig.~\ref{fig:mitchelllog2} and the corresponding exponentiation require a significantly smaller area than using lookup tables and allow significantly higher clock frequencies. As the detection results are similar for both approaches, as seen from Figs.~\ref{fig:ROCMULtable} and \ref{fig:ROCMULMitchell}, only the Mitchell approximation version is considered in the following.

\subsection{Sub-Expression Sharing}
As mentioned in Section~\ref{sec:proposedarchitecture}, the number of adders for performing the matrix-vector multiplications by $\mathbf{A}$ and $\mathbf{A}^T$ can be reduced.
Here, we use an iterative two-term sub-expression sharing approach with the constraint that each addition should be performed at the minimum depth to keep the power consumption low.
The results of this are shown in Table~\ref{tab:ses}.
As seen, a significant reduction is obtained compared to implementing the additions straightforwardly.
For comparison, the other computations for the Fast approach are, as can be seen from Fig.~\ref{fig:architecture}, $3K + \tau_{\text{p}}T = 3200$ additions/subtractions and $2K$ multiplications, of which $K$ are reduced to a simple shift, so $K=1024$ multiplications. For the Mult approach, $2K = 2048$ additions/subtractions, $K=1024$ logarithm approximations, and $K=1024$ exponent approximations are used, as evident from Fig.~\ref{fig:MULarchitecture}. 
\begin{table}
	\centering
	\caption{Number of Adders for the Matrix-Vector Multiplications without and with Sub-Expression Sharing in the Implemented Example.\label{tab:ses}}
	\begin{tabular}{c|cc|c}
		& \multicolumn{2}{c|}{Adders} & \\
		Matrix & Without & With & Reduction\\
		\hline
		$\mathbf{A}$ & 8064 & 5699 & 29.3\%\\
		$\mathbf{A}^T$ & 7168 & 4709 & 34.3\%\\
		\hline 
		Total & 15232 & 10408 & 31.7\%
	\end{tabular}
\end{table}

\begin{figure}
	\centering
\subfigure[]{\label{fig:aarea}\includegraphics[scale=0.3]{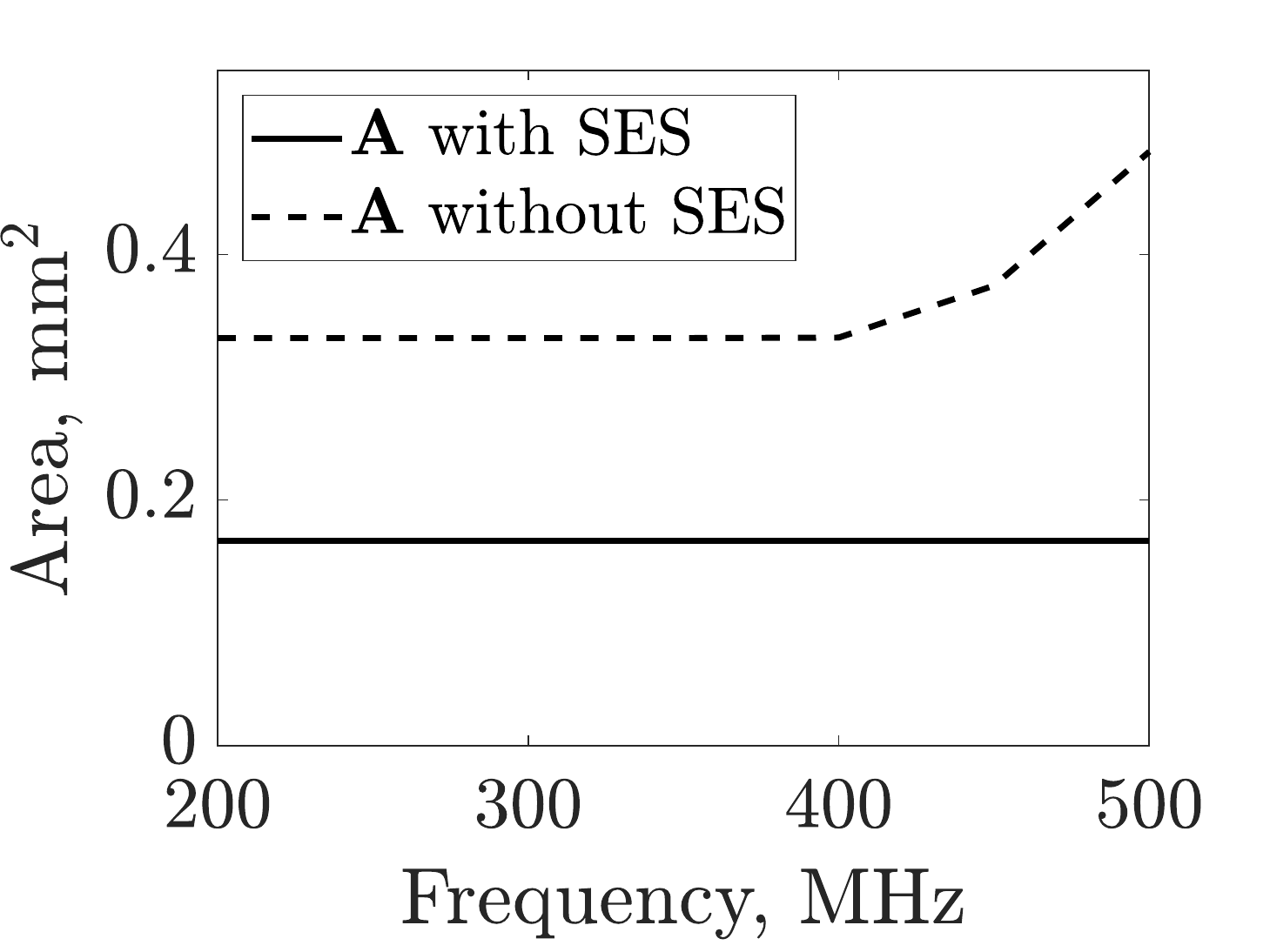}}
	\hspace{-3mm} 
\subfigure[]{\label{fig:apower}\includegraphics[scale=.3]{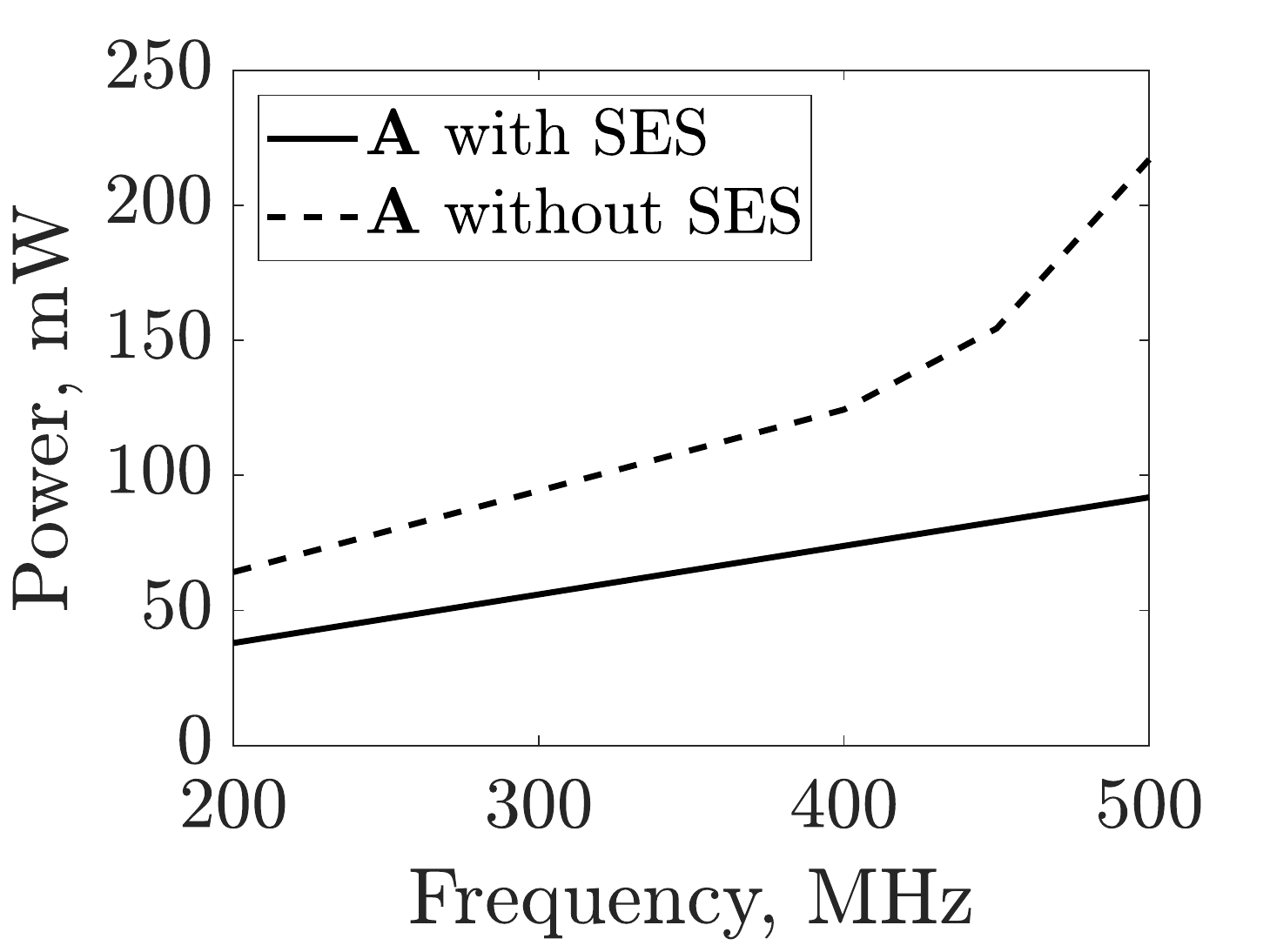}}  \caption{ (a) Area  and (b) power results at different clock frequencies for the matrix-vector multiplication with $\mathbf{A}$ with and without using sub-expression sharing.}
\label{fig:apa}	
\end{figure}

\begin{figure}
	\centering
	\subfigure[]{\label{fig:atarea}\includegraphics[scale=0.3]{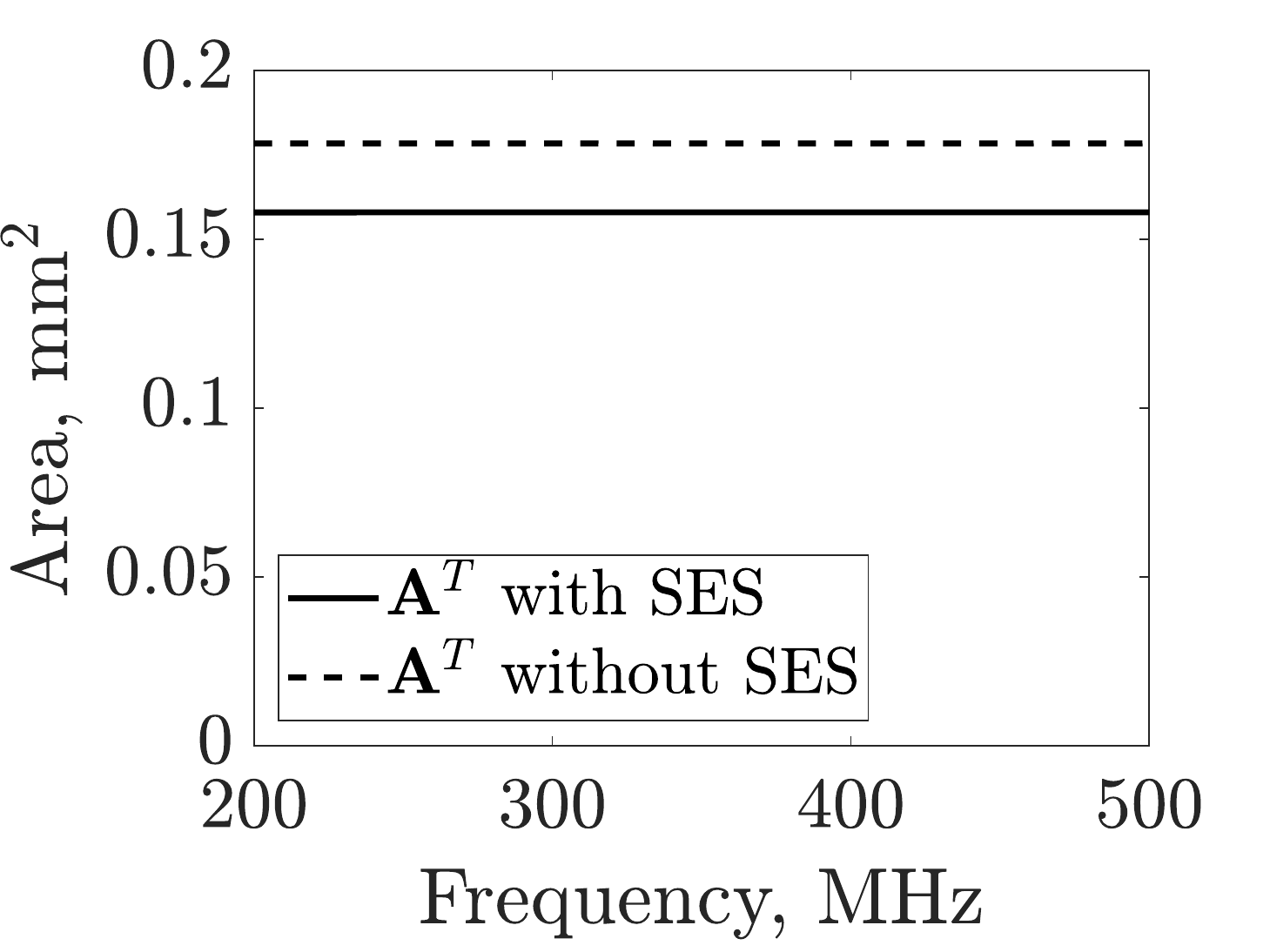}}
		\hspace{-3mm} 
	\subfigure[]{\label{fig:atpower}\includegraphics[scale=.3]{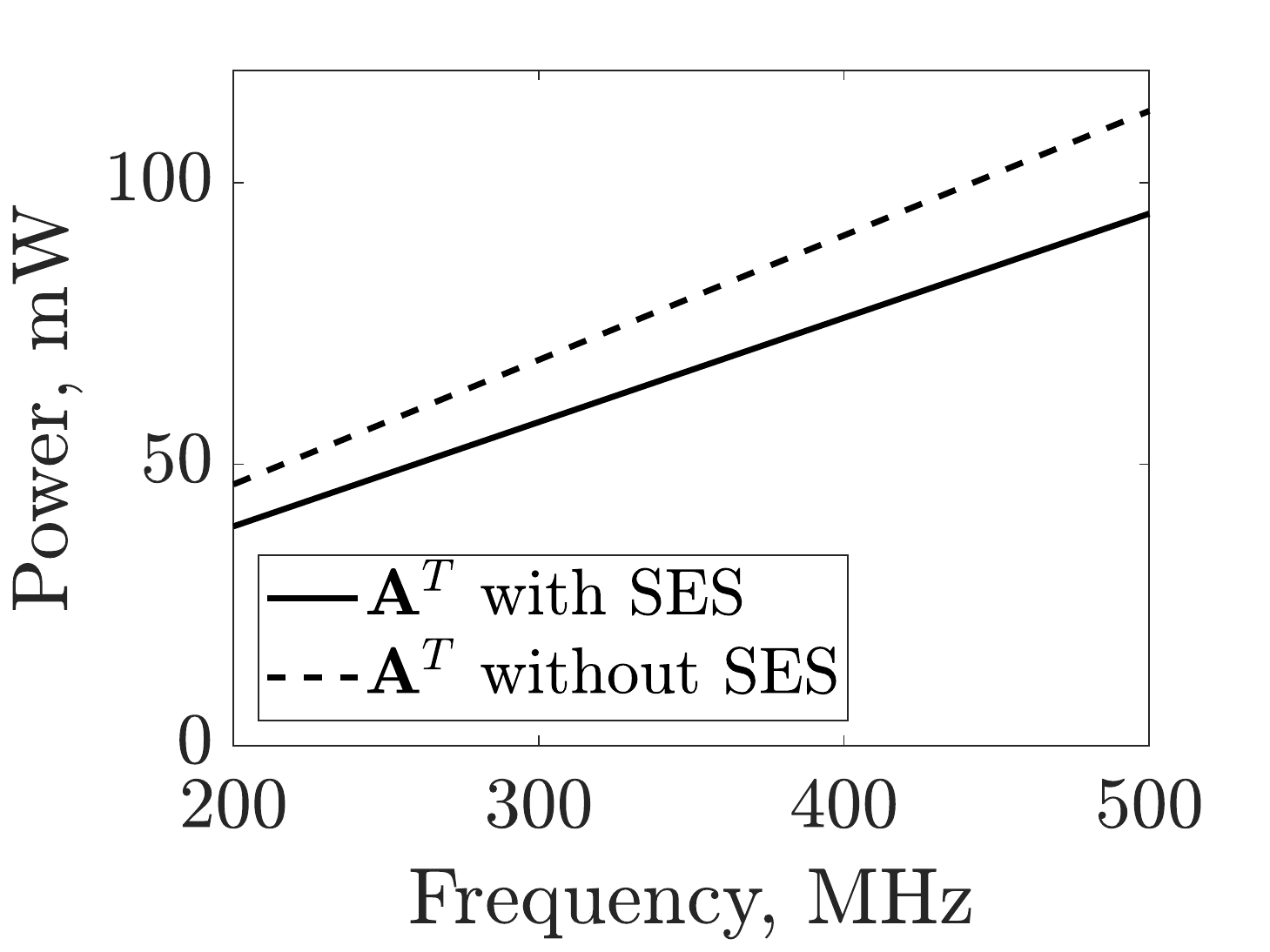}}  \caption{ (a) Area  and (b) power consumption at different clock frequencies for the matrix-vector multiplication with $\mathbf{A}^T$ with and without using sub-expression sharing.}
	\label{fig:apat}	
\end{figure}

To further clarify the consequences of sub-expression sharing, the $\mathbf{A}$ and $\mathbf{A}^T$ matrix-vector multiplications have been synthesized separately for using the word lengths of the Fast implementation, i.e., the word lengths in Fig.~\ref{fig:wl}. The area and power results are shown in Figs.~\ref{fig:apa}~and~\ref{fig:apat} for $\mathbf{A}$ and $\mathbf{A}^T$, respectively. It can be seen that although the reduction in area is not the $\approx 30\%$ expected, more for $\mathbf{A}$ and less for $\mathbf{A}^T$, sub-expression sharing provides a significant area and power reduction.

\subsubsection{Selecting $\mathbf{A}$}

The number of ones in $\mathbf{A}$ is always $KT$, as each user uses $T$ pilots. This leads to that the number of adders without sharing is the same independent of the pilot hopping sequences: $KT-\tau_{\text{p}}T$ for $\mathbf{A}$ and $KT-K$ for $\mathbf{A}^T$.
However, by selecting different pilot hopping sequences, different sub-expression sharing results can be obtained.
Therefore, one can try to optimize $\mathbf{A}$ to give fewer adders, while still having good detection performance.
One may note that qualitatively, detection performance increases the more different the pilot hopping sequences are, while the sharing increases with similarities in pilot hopping sequences.

Similarly, one can see that decreasing the length of the pilot hopping sequences $T$ will lead to fewer adders, assuming that the number of users is constant. 
In \cite{Becirovic2019}, it was shown that detection performance scales with $\tau_{\text{p}}T$.
Hence, to decrease the implementation complexity with similar detection performance, one may decrease $T$ and increase $\tau_{\text{p}}$.
However, this will lead to that less time is available to transmit data.

\subsection{Implementation Results\label{sec:results}}
\begin{figure}
	\centering
	\includegraphics[scale=.38]{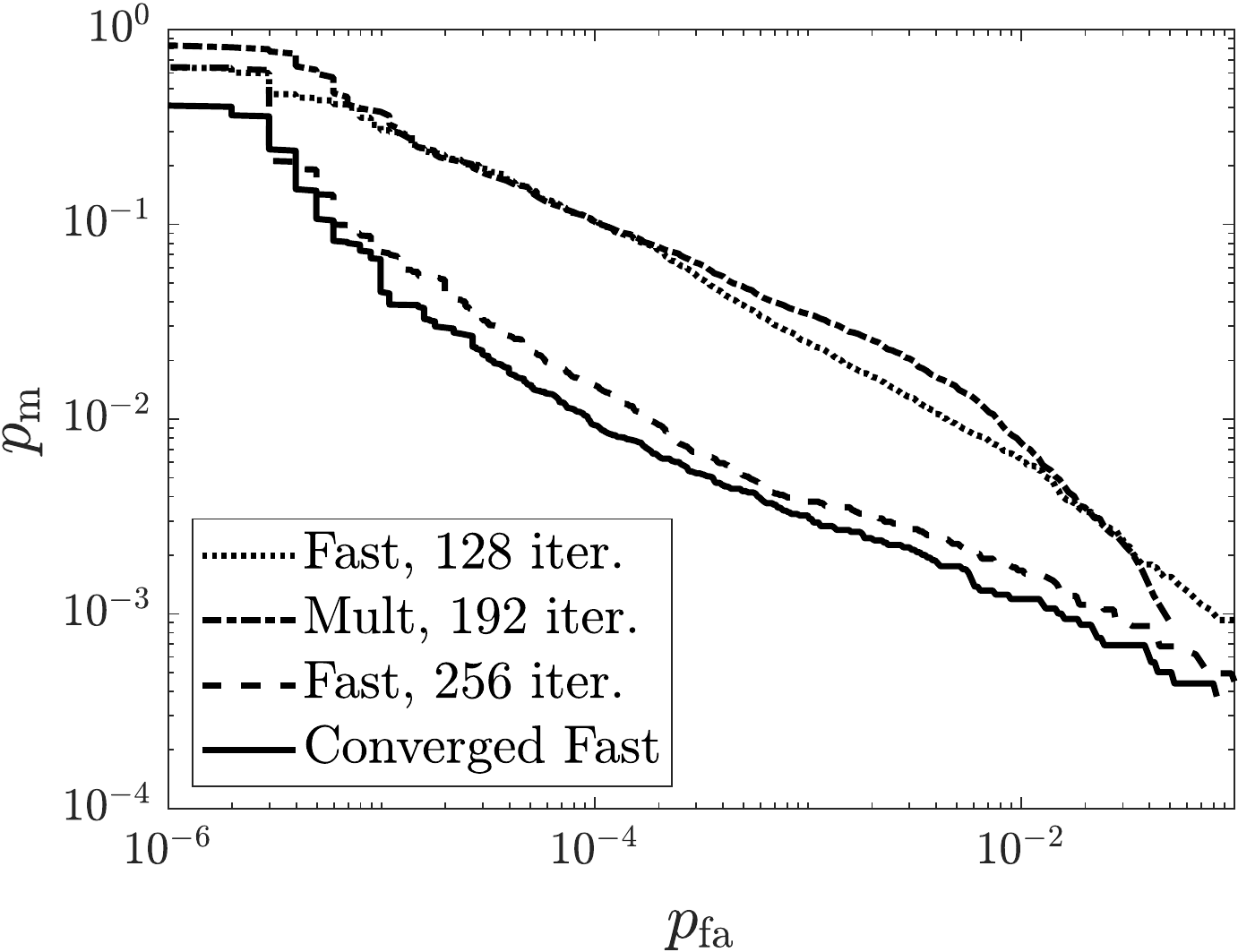}
	\caption{Missed detection, $p_{\text{m}}$, versus false alarm rates, $p_{\text{fa}}$, of the three considered implementation examples and the converged floating-point results.}
	\label{fig:implementedROC}
\end{figure}

Based on the previous analyses, three different design cases are considered. A design using the Fast algorithm and 256 iterations provides detection results close to the converged floating-point results as shown in Fig.~\ref{fig:ROCfast}. For the Mult algorithm, 192 iterations are used since increasing the number of iterations beyond that only gives a limited detection gain. As discussed earlier, the Mult algorithm converges faster for the good cases, but slower for the bad cases, so depending on channel conditions etc, the actual detection performance may differ, and the different approaches may be beneficial in different scenarios. Finally, the Fast algorithm with 128 iterations gives about the same detection performance as the Mult algorithm with 192 iterations and is also included. The design is identical, and it is only the number of iterations in the power simulations that differ. The corresponding ROC curves are shown in Fig.~\ref{fig:implementedROC}.

\begin{figure}
	\centering
	\subfigure[]{\label{fig:aFastMult}\includegraphics[scale=0.3]{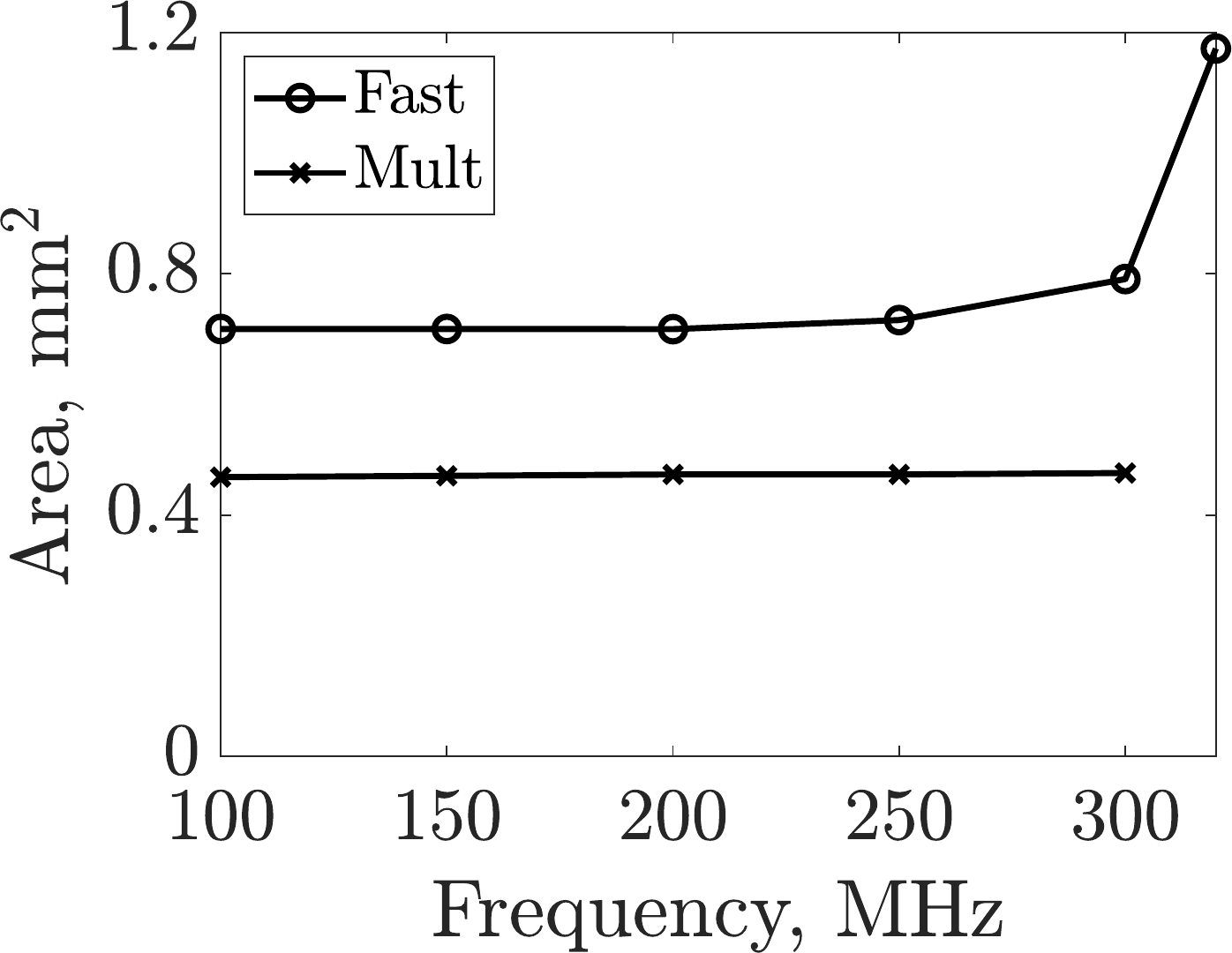}}
	\hfill
	\subfigure[]{\label{fig:pFastMult}\includegraphics[scale=.3]{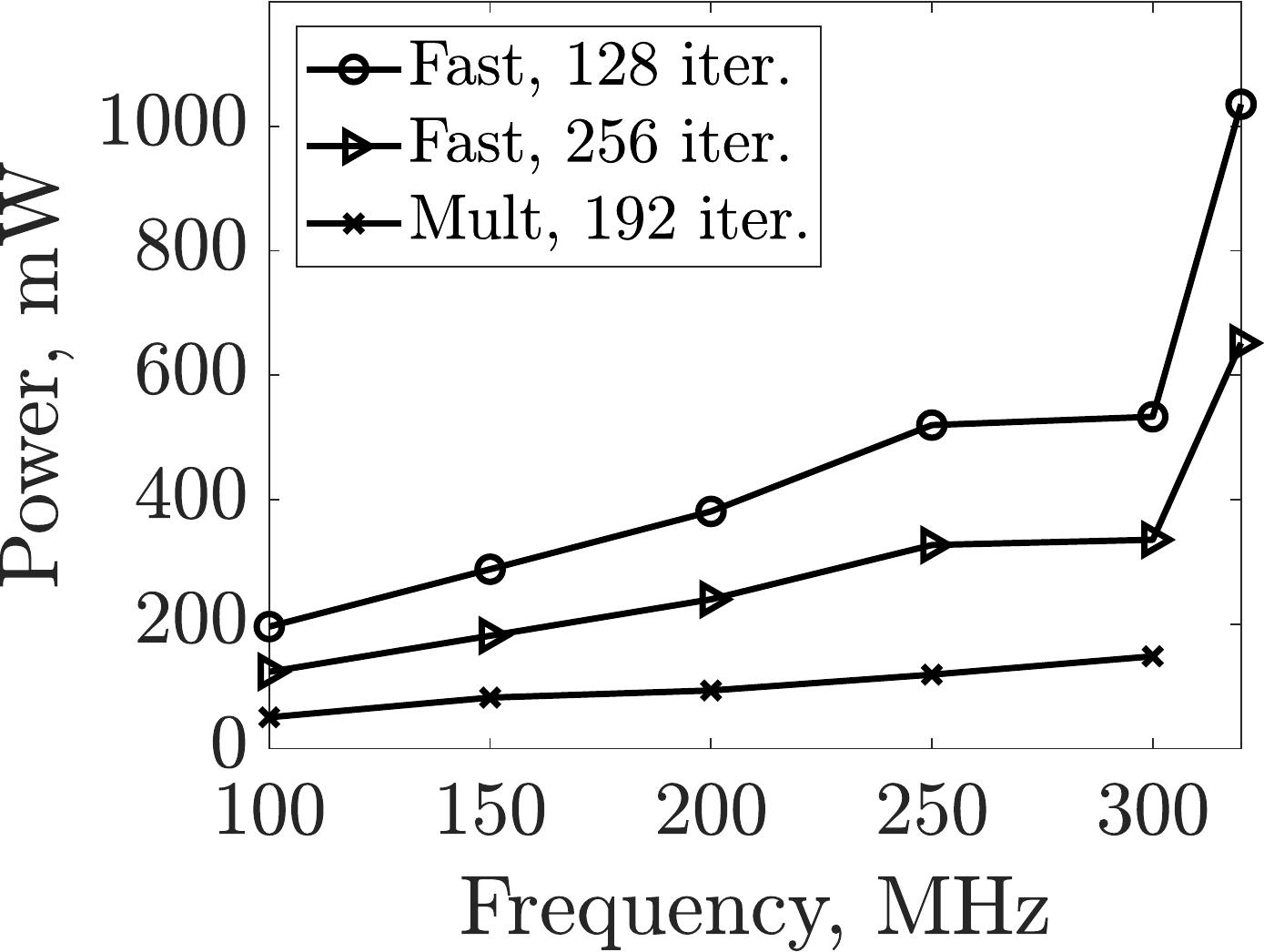}}  \caption{a) Area and (b) power consumption at different clock frequencies for the implementations of Fast and Mult algorithms. }
	\label{fig:apFastMult}
\end{figure}

The designs were synthesized for increasing clock frequency and power simulated. The results in terms of area and power are shown in Fig.~\ref{fig:apFastMult}. It can be seen that the area and power is significantly smaller for the Mult algorithm, although the maximum clock frequency is slightly higher for the Fast algorithm, 320~MHz, as compared to 300~MHz for the Mult algorithm\footnote{Can be increased to 320~MHz by using a different programmable shift with a small area increase.}.

It may at first seem a bit unintuitive that the power consumption for the Fast approach using 128~iterations is higher than when using 256 iterations. However, it should be noted that these results are against clock frequency, so the approach with 128~iterations can perform twice as many detections. Also, as illustrated in the following, the switching activity and therefore power consumption is reduced when the results converge. Similarly, the Mult approach using 192~iterations can perform more detections compared to the Fast approach with 256~iterations, but fewer than the Fast approach with 128~iterations.

In the following, the 250~MHz clock frequency designs are used. Considering the area curves in Fig.~\ref{fig:apFastMult}, this should be close to the minimum area design, as synthesizing for a lower clock rate will not decrease the area significantly.

\begin{figure}
	\centering
	\subfigure[]{\label{fig:blockpower}\includegraphics[scale=.3]{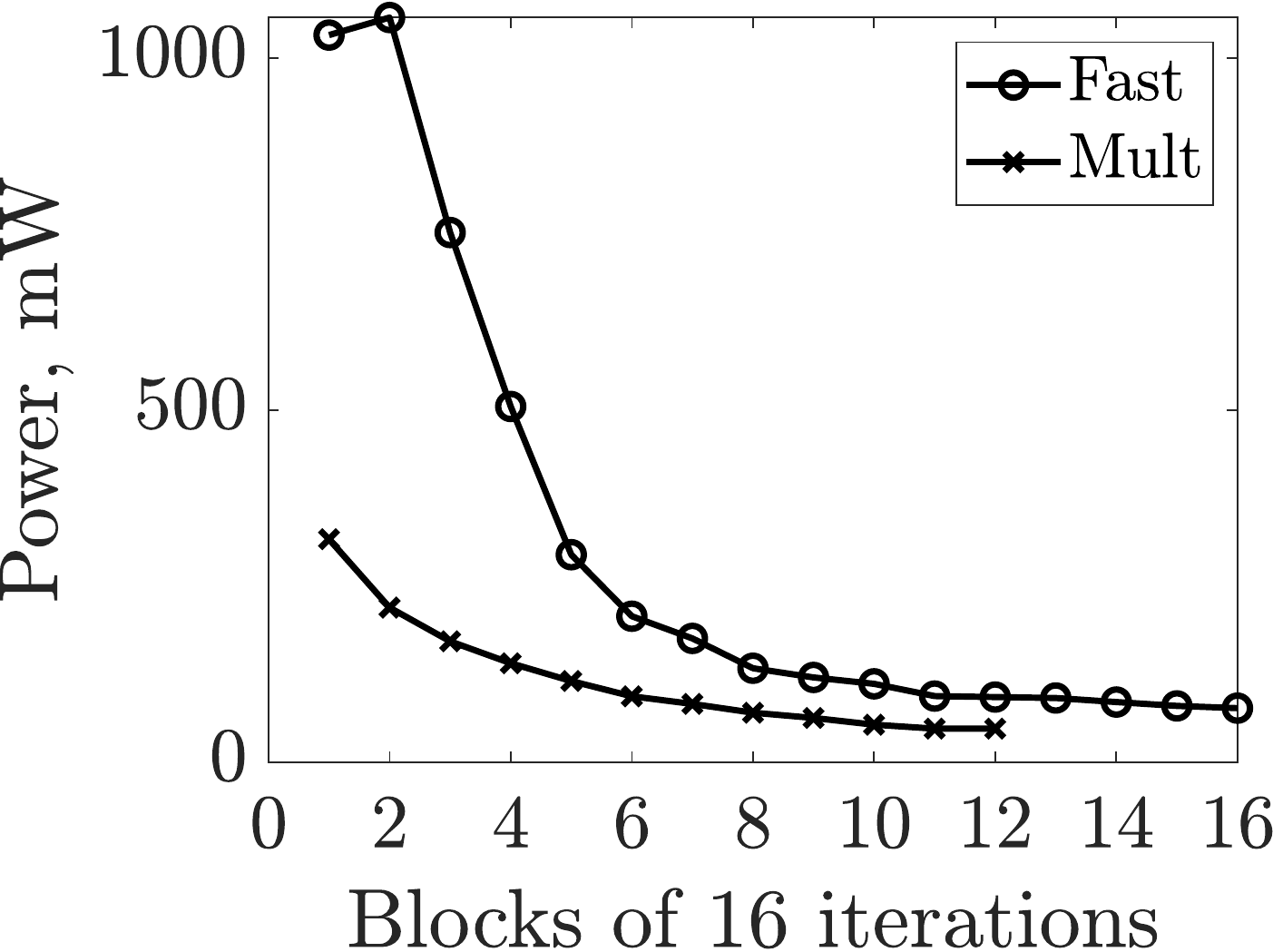}}
	\hfill
	\subfigure[]{\label{fig:energy}\includegraphics[scale=.3]{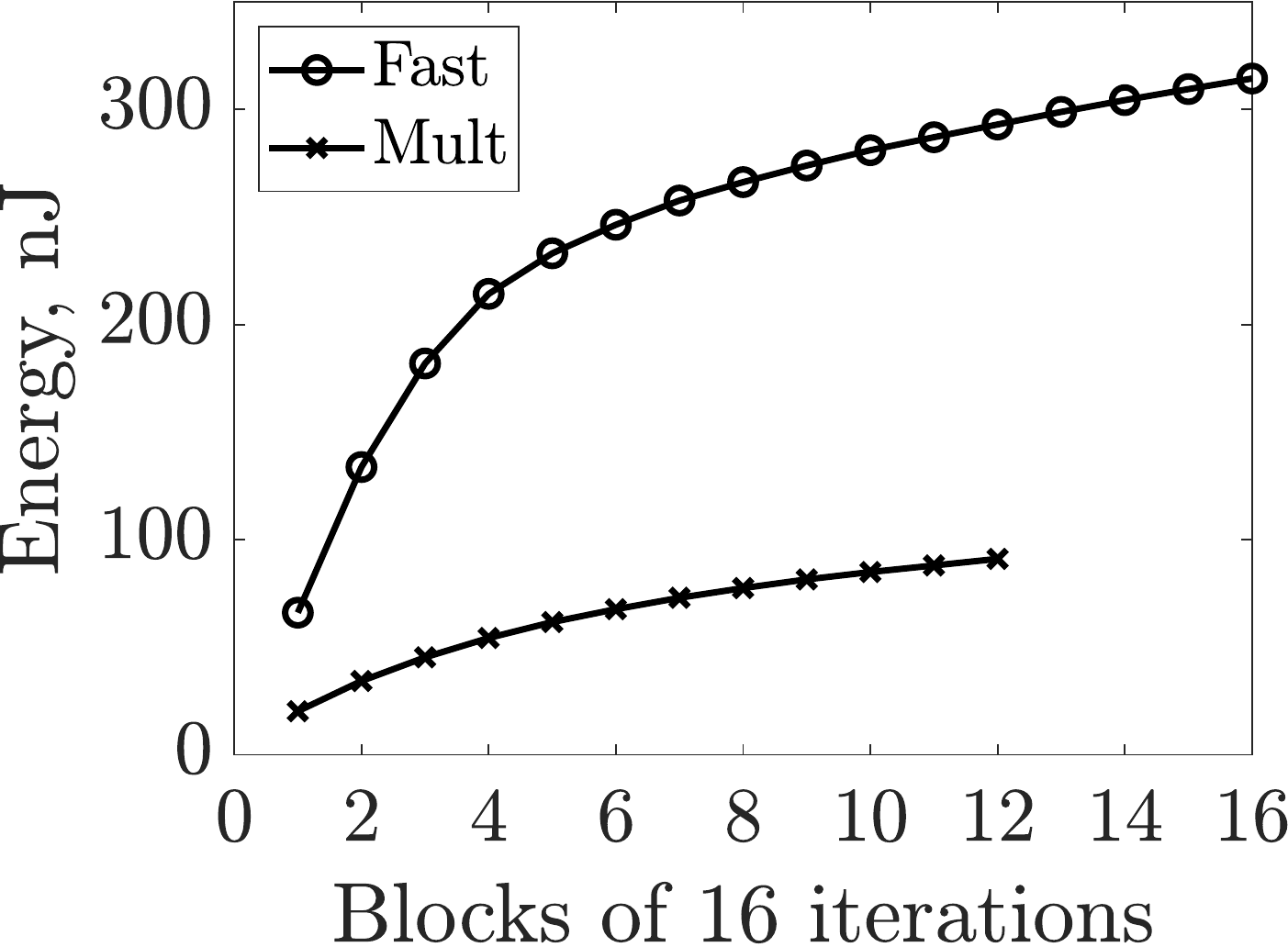}}
	\caption{(a) Average power and (b) accumulated energy consumption over blocks of 16 iterations for Fast and Mult approaches. }
	\label{fig:blockpowerenergy}
\end{figure}

To provide further insights here, the average power consumption was analyzed in blocks of 16 iterations. The results are shown in Fig.~\ref{fig:blockpower}. As expected, the power consumption is high in the first iterations and decreases as the results converge.

Similarly, one can compute the energy required for one detection instance. This is shown in Fig.~\ref{fig:energy}, where the 16 iteration block averages are used as underlying data. Here, it is clear that the total energy required for one detection is significantly lower for the Mult algorithm compared to the Fast algorithm. Even if 128~iterations are used for the Fast algorithm, about 250~nJ is required, compared to about 100~nJ for the Mult algorithm. Still, the Fast algorithm has better detection performance when more iterations are performed.

Finally, it should be pointed out that the current approaches can perform about one million detections per second. This is most likely too much in the current settings. However, it is possible that the future brings scenarios with higher detection rate requirements. Hence, in the meanwhile one can reduce the energy consumption further by power supply voltage scaling, leading to even lower power and energy consumption.

\subsection{Comparison with Previous Results}
As discussed earlier, to the best of the authors' knowledge, the only other existing previous implementations of mMTC activity detection algorithms are \cite{Tran2019,Henriksson2020}. However, this is for the other type of scenario, non-orthogonal sequences, instead of the pilot-hopping scheme considered here. Also, \cite{Tran2019} considers the case where each user can send different sequences and therefore transmit a few bits of information (four bits). Hence, the comparison is not fully relevant as such.  The reported detection performance is better for both \cite{Tran2019} and \cite{Henriksson2020}. However, this will apart from the algorithm also depend on, e.g., SNR, sequence length, number of users, user activity etc.

The implementations in \cite{Tran2019} and \cite{Henriksson2020} can perform about 11600 and 289 detections per second, respectively, so significantly less than the proposed architectures. The implementation occupies about 5.1~mm$^2$ and 1.14~mm$^2$ chip area for \cite{Tran2019} and \cite{Henriksson2020}, respectively in the same standard cell process, memories not included. This should be compared to about 0.7~mm$^2$ and 0.4~mm$^2$ for the Fast and Mult algorithms, respectively. Hence, both the proposed architectures occupies significantly less chip area.

The energy consumption for one detection is about 0.13~mJ and 4.5~mJ for \cite{Tran2019} and \cite{Henriksson2020}, respectively. The Fast algorithm presented here requires about 250~nJ and 300~nJ for 128 and 256 iterations, respectively. The Mult algorithm requires about 100~nJ. Hence, the proposed approaches are several orders of magnitude more energy efficient.

Another advantage of the approaches in \cite{Tran2019,Henriksson2020} is that the pilot sequences are easy to change. In the approaches proposed in the current work, they are hard coded into the structure of the matrix-vector multiplications with $\mathbf{A}$ and $\mathbf{A}^T$. While this provides a very efficient implementation, it may have other drawbacks. However, one may imagine that a system comes with prepared user modules. This will for example guarantee that all users have unique pilot-hopping sequences. One can also consider to synthesize the proposed architectures to FPGAs, in which case the sequences are readily configurable through a re-synthesis of the $\textbf{A}$ and $\textbf{A}^T$ matrix-vector multiplications\footnote{The spectral radius is about the same for all matrices with the same parameters that have been tested.}.

\section{Conclusion}
\label{sec:conclusions}
Two different architectures for activity detection in grant-free random access massive machine type communication based on pilot-hopping sequences have been proposed. They both solve a non-negative least squares problem, but with different algorithms. Both algorithms are iterative and with deterministic computations, enabling a high-speed implementation. It is shown that the fixed-point implementation of the fast projected gradient algorithm, Fast, provides detection results close to the converged floating-point realization. The multiplicative updates algorithm, Mult, is shown to converge faster for good cases, but slower for bad cases. It is partially implemented in the logarithmic domain to replace multiplications and divisions with additions and subtractions. Here, 192 iterations are selected, which gives a detection performance similar to 128 iterations of the Fast algorithm. However, the area and power/energy consumptions of the multiplicative updates algorithm is significantly lower. The energy consumed for one detection is about 100, 250, and 300 nJ for Mult with 192 iterations, Fast with 128 iterations, and Fast with 256 iterations, respectively, in a scenario with 1024 users and eight pilots, each with a length of 16.

The detection rate is more than one million detections per second, which is most likely too much for contemporary scenarios. However, future scenarios will most likely require an increased detection speed. It is also possible to reduce the clock frequency and power supply voltage to benefit from a lower energy consumption. Currently, the pilot hopping matrices are hard coded in the structure, requiring that the terminals must send the correct sequences. When implementing the same architecture on an FPGA, it is possible to change the pilot hopping matrices by reprogramming the FPGA. This is readily obtained as the architectures have code generators developed.

\ifCLASSOPTIONcaptionsoff
  \newpage
\fi



\bibliographystyle{IEEEtran}
\bibliography{IEEEabrv,references}
%

%

\begin{IEEEbiography}[{\includegraphics[width=1in,height=1.25in,clip,keepaspectratio]{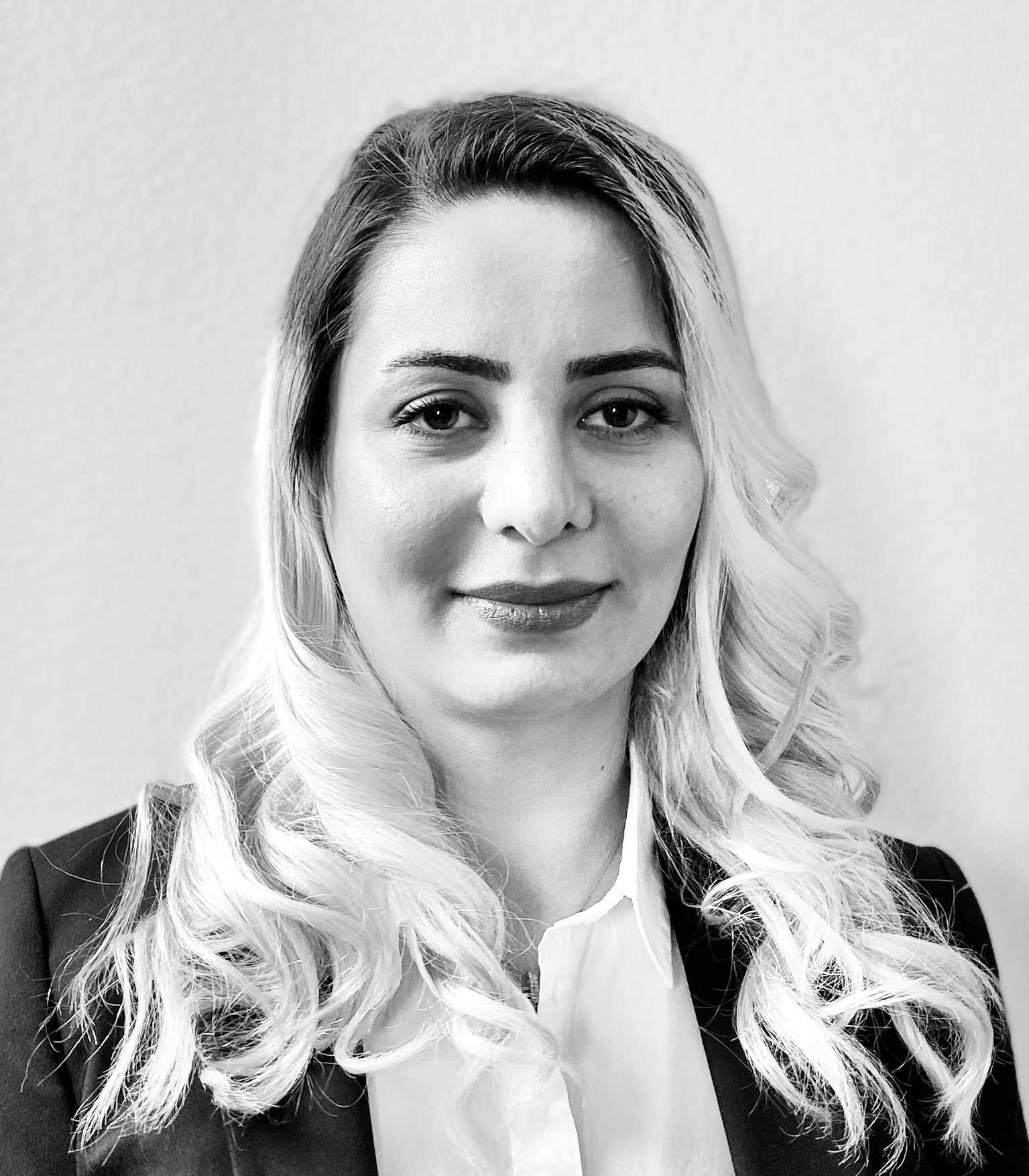}}]{Narges Mohammadi Sarband} (S'18)
received her B.Sc. degree in computer hardware engineering from Sadjad University of Technology, Mashhad, Iran in 2006 and her M.Sc. degree in computer architecture from  University of Isfahan, Isfahan, Iran in 2014. She is from 2017 working towards her Ph.D. in electrical engineering with specialization in computer engineering  at Linköping University (LiU), Sweden.
Her research focuses on efficient design and hardware implementation of communication algorithms for beyond 5G systems in both ASIC and FPGA. She has also worked on optimizing peer to peer algorithms in computer networking.
\end{IEEEbiography}

\begin{IEEEbiography}[{\includegraphics[width=1in,height=1.25in,clip,keepaspectratio]{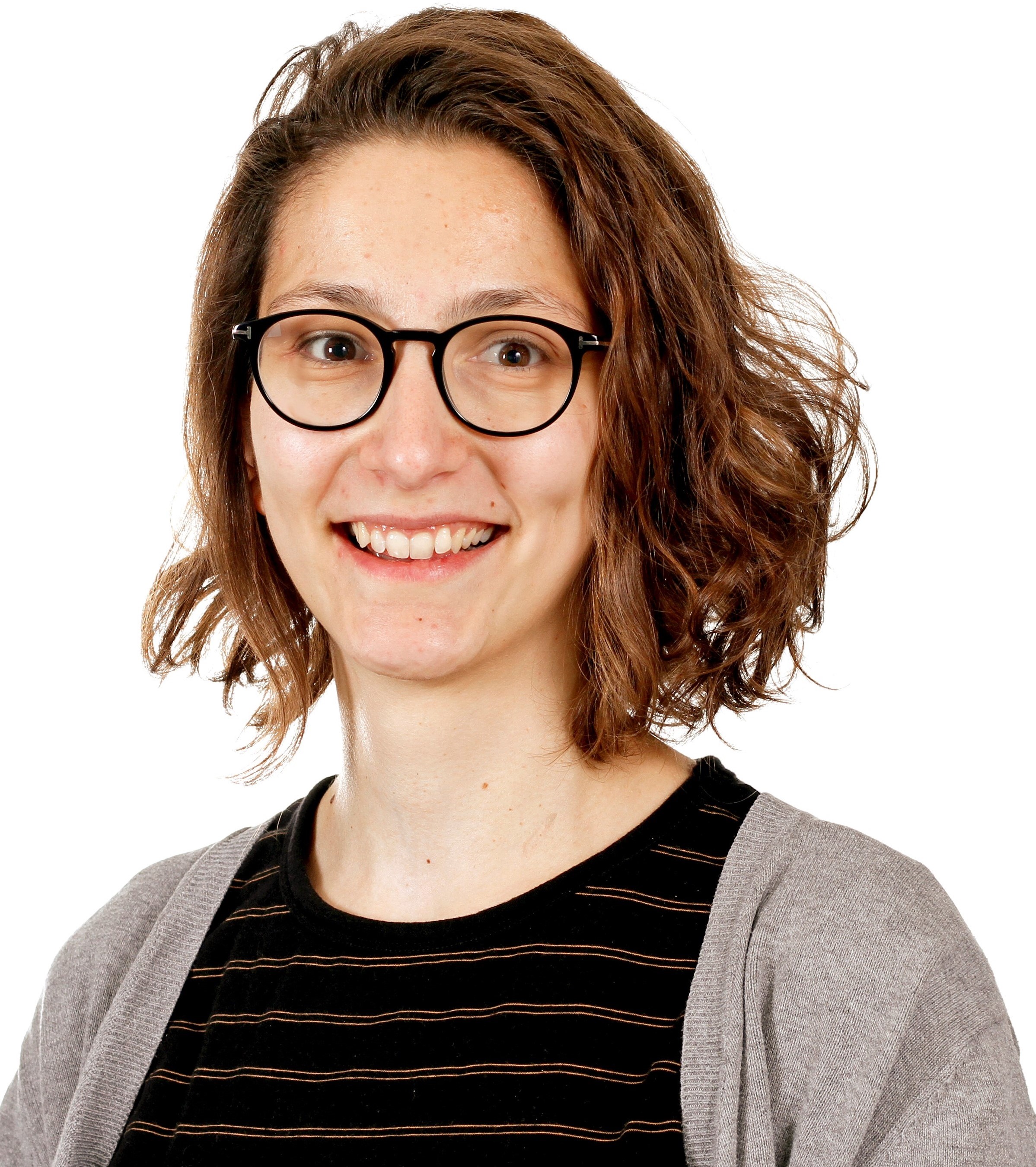}}]{Ema Becirovic}
	 (GSM'18) received the M.Sc. degree from Linköping University, Sweden, in 2018, where she is currently pursuing the Ph.D. degree with the Department of Electrical Engineering, Division of Communication Systems. Her research is mainly focused on massive MIMO and massive machine-type communications.
\end{IEEEbiography}

\begin{IEEEbiographynophoto}{Mattias Krysander}
	is an associate professor in the De-partment  of  Electrical  Engineering,  Linköping  University, Sweden.  His  research  interests  include  model-based  and  data-driven  diagnosis  and  prognosis.  To  address  the  complexity  and  size  of  industrial  systems  (mainly  vehicle  systems), he has used structural representations of models and developed graph theoretical methods for assisting the design of diagnosis systems and fault isolation and sensor placement analysis.
\end{IEEEbiographynophoto}

\begin{IEEEbiography}[{\includegraphics[width=1in,height=1.25in,clip,keepaspectratio]{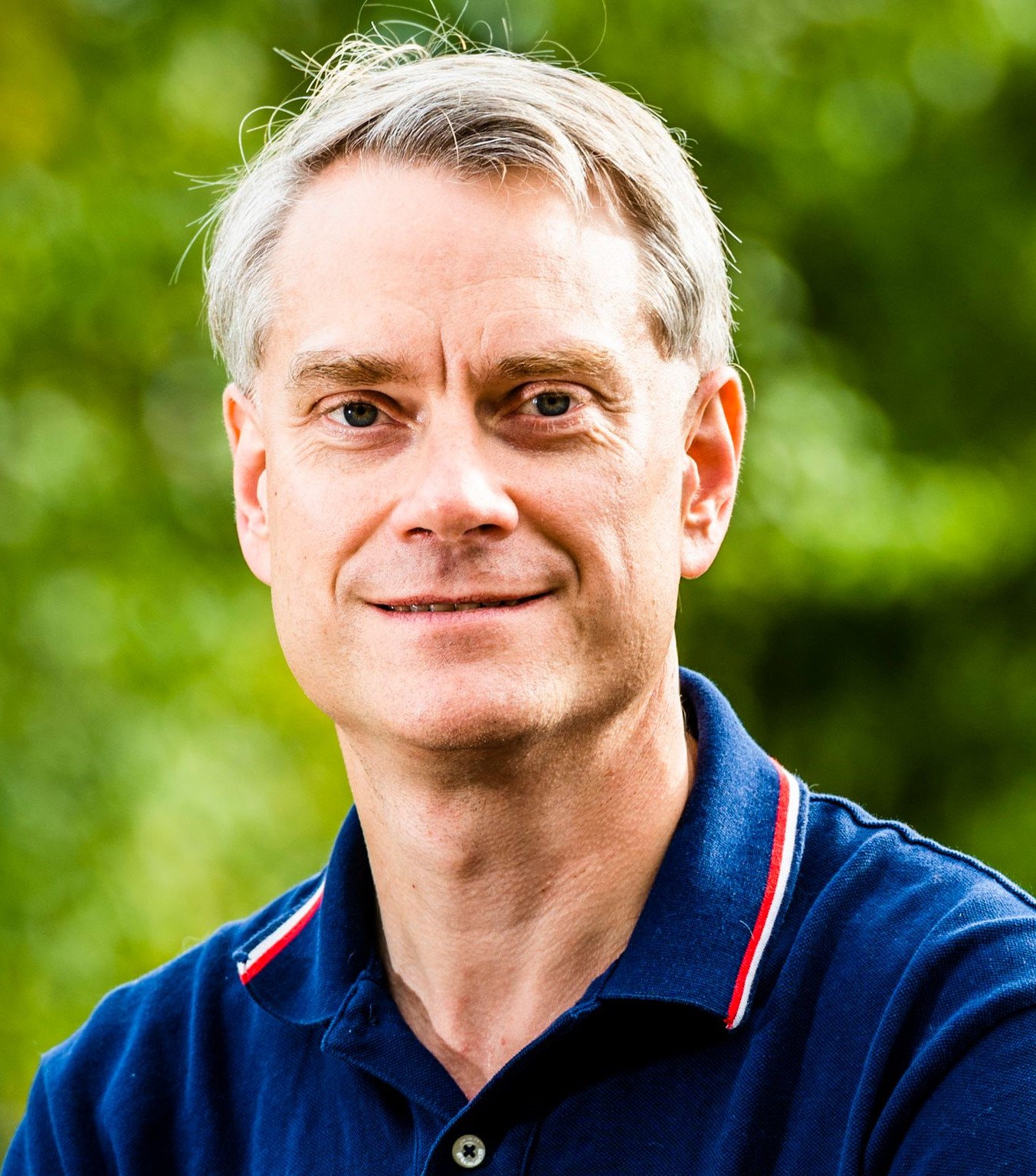}}]{Erik G. Larsson} (S'99--M'03--SM'10--F'16)
	received the Ph.D. degree from Uppsala University,
	Uppsala, Sweden, in 2002.  He is currently Professor of Communication
	Systems at Link\"oping University (LiU) in Link\"oping, Sweden. He was
	with the KTH Royal Institute of Technology in Stockholm, Sweden, the
	George Washington University, USA, the University of Florida, USA, and
	Ericsson Research, Sweden.  His main professional interests are within
	the areas of wireless communications and signal processing. He 
	co-authored \emph{Space-Time Block Coding for  Wireless Communications} (Cambridge University Press, 2003) 
	and \emph{Fundamentals of Massive MIMO} (Cambridge University Press, 2016). 
	He is co-inventor of 19 issued U.S. patents.
	
	Currently he is an editorial board member of the \emph{IEEE Signal
		Processing Magazine}, and a member of the  \emph{IEEE Transactions on Wireless Communications}    steering committee. 
	He served as  chair  of the IEEE Signal Processing Society SPCOM technical committee (2015--2016), 
	chair of  the \emph{IEEE Wireless  Communications Letters} steering committee (2014--2015), 
	General respectively Technical Chair of the Asilomar SSC conference (2015, 2012), 
	technical co-chair of the IEEE Communication Theory Workshop (2019), 
	and   member of the  IEEE Signal Processing Society Awards Board (2017--2019).
	He was Associate Editor for, among others, the \emph{IEEE Transactions on Communications} (2010-2014) 
	and the \emph{IEEE Transactions on Signal Processing} (2006-2010).
	
	He received the IEEE Signal Processing Magazine Best Column Award
	twice, in 2012 and 2014, the IEEE ComSoc Stephen O. Rice Prize in
	Communications Theory in 2015, the IEEE ComSoc Leonard G. Abraham
	Prize in 2017, the IEEE ComSoc Best Tutorial Paper Award in 2018, and
	the IEEE ComSoc Fred W. Ellersick Prize in 2019.
	
	He is a Fellow of the IEEE.
\end{IEEEbiography}

	\begin{IEEEbiography}[{\includegraphics[width=1in,height=1.25in,clip,keepaspectratio]{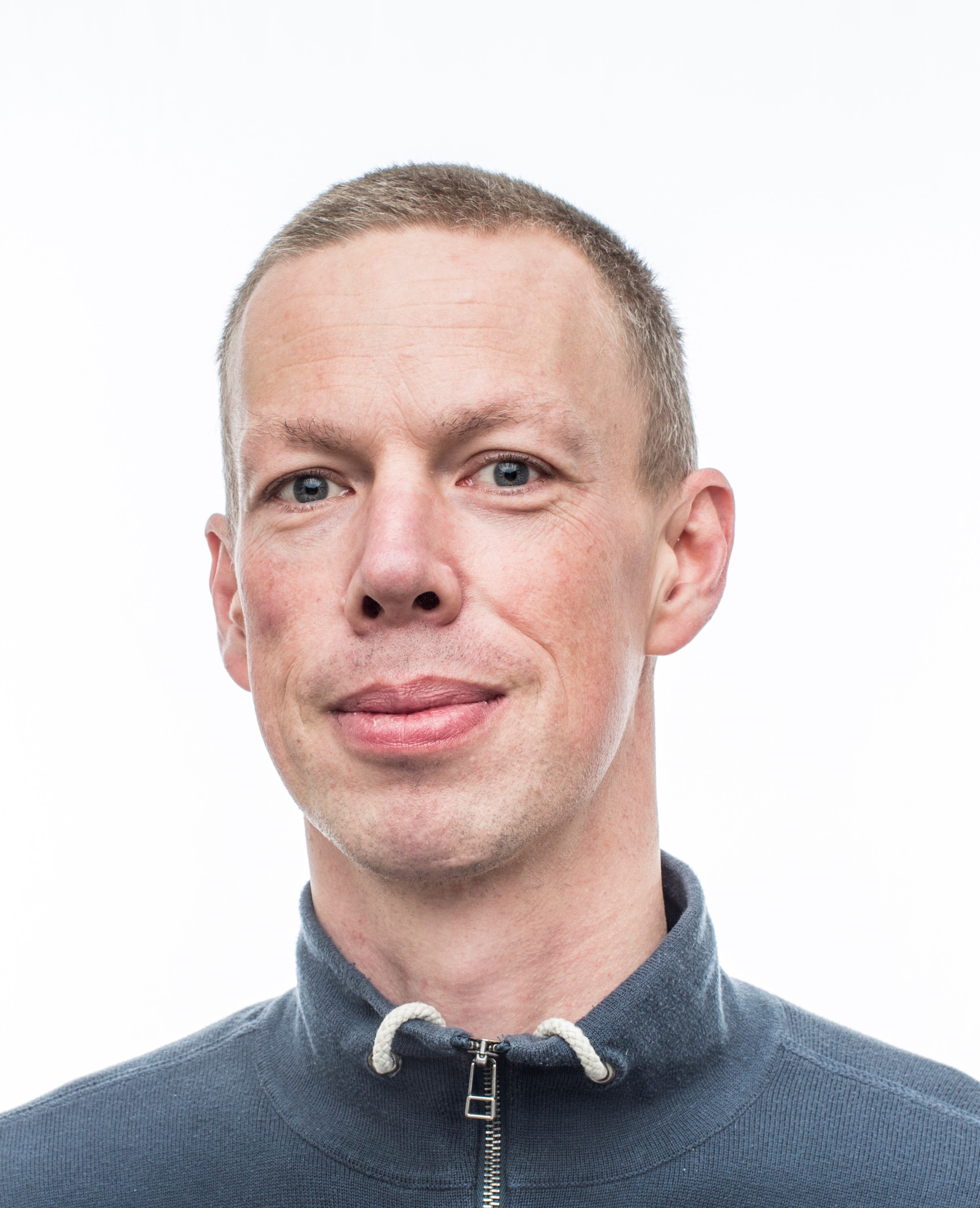}}]{Oscar Gustafsson} (S'98--M'03--SM'10)
 received the M.Sc., Ph.D., and Docent degrees from  Linköping University, Linköping, Sweden, in 1998, 2003, and 2008, respectively. He is currently an Associate Professor and Head of the  Division of Computer Engineering, Department of Electrical Engineering,  Linköping University. 
His research interests include design and implementation of DSP algorithms and arithmetic circuits in FPGA and ASIC. He has authored and co-authored over 180 papers in international journals and conferences on these topics. Dr. Gustafsson is a member of the VLSI Systems and Applications and the Digital Signal Processing technical committees of the IEEE Circuits and Systems Society. He has served as an Associate Editor for the IEEE Transactions on Circuits and Systems Part II: Express Briefs and Integration, the VLSI Journal. Furthermore, he has served and serves in various positions for conferences such as ISCAS, PATMOS, PrimeAsia, ARITH, Asilomar, NorCAS, ECCTD, and ICECS.
\end{IEEEbiography}





\end{document}